\documentclass[notitlepage]{article}

\usepackage{amssymb}
\usepackage{amsmath}
\usepackage[twoside]{geometry}
\usepackage[dvips]{graphicx}

\input{tcilatex}

\begin{document}

\date{}
\title{\textbf{Modular application of an Integration by Fractional Expansion
(IBFE) method to multiloop Feynman diagrams}}
\author{Iv\'{a}n Gonz\'{a}lez\thanks{%
e-mail: igonzalez@fis.puc.cl} \\
Departamento de F\'{\i}sica\\
Pontificia Universidad Cat\'{o}lica de Chile\\
Santiago, Chile \and Iv\'{a}n Schmidt\thanks{%
e-mail: ivan.schmidt@usm.cl} \\
Departamento de F\'{\i}sica y Centro de Estudios Subat\'{o}micos \\
Universidad T\'{e}cnica Federico Santa Mar\'{\i}a\\
Valparaiso, Chile}
\maketitle

\begin{abstract}
We present an alternative technique for evaluating multiloop Feynman
diagrams, using the integration by fractional expansion method. Here we
consider generic diagrams that contain propagators with radiative
corrections which topologically correspond to recursive constructions of
bubble type diagrams. The main idea is to reduce these subgraphs, replacing
them by their equivalent multiregion expansion. One of the main advantages
of this integration technique is that it allows to reduce massive cases with
the same degree of difficulty as in the massless case.
\end{abstract}


\bigskip


\vfill\newpage

\section{Introduction}

\qquad The evaluation of multiloop Feynman diagrams is currently one of the
most important problems in Quantum Field Theory. Thanks to the development
of analytical techniques in high order perturbative calculations, it has
become possible to compare precision experimental measurements with the
theoretical models that try to explain them. There are several of these
techniques, and some of the best known can even be found in textbooks \cite%
{AGr, VSm}. One that is not particularly used but which nevertheless has
advantages for the evaluation of some complicated Feynman diagrams is called
Negative Dimension Integration Method (NDIM). The basic foundations of NDIM
were initially suggested in the work of Halliday and Ricotta \cite{IRi},
which using the dimensional regularization prescription $\left(
D=4-2\epsilon \right) $\ make an analytical continuation of the dimension $D$
into negative values, something which can be done since the Feynman
integrals are in fact analytic in arbitrary dimension $D$. In a previous
work \cite{IGoIBFE} we proposed that a more appropriate name for this
technique should be integration by fractional expansion (IBFE), representing
better its mathematical and physical basis.

The purpose of the IBFE technique is to transform the Schwinger integral
parametric representation of a specific Feynman diagram into an equivalent
mathematical structure, which contains several summations and Kronecker
deltas and which we call multiregion expansion (MRE) of the diagram. This
name comes from the fact that this particular expansion is made around the
values zero and infinity simultaneously, and that once the expression is
summed using the Kronecker deltas, these expansions become explicitly
separated. In general the number of summations is bigger than the number of
deltas that are generated, and in this case there are several different ways
of summing using the Kronecker deltas, in fact $C_{\delta }^{\sigma }=\dfrac{%
\sigma !}{\delta !\left( \sigma -\delta \right) !}$ different forms, where $%
\delta $ is the number of Kronecker deltas and $\sigma $ the number of sums.
Each one of these forms give rises in general to generalized and
multivariable hypergeometric functions.

Several authors have used and formalized this method, with a wide variety of
applications to mainly one loop \cite{CAn2,ASu11,ASu13,ASu14,CAn1} and two
loop \cite{ASu2,ASu3,ASu4,ASu6,ASu10} diagrams. In a previous work we
presented a optimization of this technique \cite{IGoIBFE}, which allows to
evaluate more complex diagrams and even certain families of $L$ loop
diagrams. In this work we will show that is possible to improve the
efficiency of the IBFE method when it is applied in a modular form to a
diagram, that is applying it iteratively to subgraphs or modules that
contain one or more contiguous loops of the same diagram. The procedure
consists in replacing the Schwinger integral parametric representation of
each module by its corresponding MRE, and simultaneously the diagram is
reduced topologically into simpler graphs. This reduction allows to get the
MRE of the complete diagram, as a product of functions that we have called $%
n $-$loop$ functions (associated to $n$ loop subgraphs), each of them being
an MRE related to one of the modules that are present in the diagram. Since
each module can have different masses distributed in different ways in its
propagators, there exist several $n$-$loop$ functions depending on the
particular configuration of masses in the propagators.

The optimization that is reached when applying the IBFE method in this way
to a diagram consists in a reduction of the obtained MRE with respect to the
one that is reached when the diagram is considered with all its loops
simultaneously. In order to explain the technique in detail, we start with
the analysis of the simplest topology that can constitute a subgraph of a
Feynman diagram: the one loop module (subgraph), the bubble diagram, which
is going to be the basis that will be used to derive $1$-$loop$ functions,
useful to evaluate a variety of Feynman diagrams.

The remainder of the paper is organized as follows. In Sec. II, we describe
how to easily obtain the Schwinger parametric representation associated to a
Feynman diagram, whose mathematical structure is the starting point for the
application of the IBFE technique. Here we present the basic elements of the
integration method, and a more rigorous extension is given in the Appendix.
In section III we briefly review the conventional way to deal with this type
of diagrams, which is loop by loop. This will provide a direct test for the
method proposed here. Later, in Sections IV and V we define and deduce the $%
1 $-$loop$ functions of the bubble module, for all the possible mass and
momentum configurations, which are a total of eight functions : $G_{k}$ and $%
\overline{G}_{k}\;\left( k=A,B,C,D\right) $. Finally, in Section VI we
develop several applications of the proposed modular procedure, which will
be compared with its equivalent MRE obtained from an application of the IBFE
technique considering all loops at the same time.

\section{Mathematical Formalism}

\qquad In what follows we will briefly describe the algebraic elements that
are needed to understand the technique.

\subsection{Schwinger's parametric representation}

\qquad The IBFE integration technique is applied directly to the Schwinger
parametric representation of a diagram, and therefore it is important to
have a simple algorithm to obtain this representation, which is what we will
do in this subsection.

Let us consider a generic topology $G$, which represents a Feynman diagram
in a scalar theory, and let us suppose that the graph has: $N$ propagators,
each one associated to the masses $\left\{ m_{1},...,m_{N}\right\} $; $L$
loops, associated to the independent internal momenta $\left\{
q_{1},...,q_{L}\right\} $); and $E$ independent external momenta $\left\{
p_{1},...,p_{E}\right\} $.

Using the dimensional regularization prescription we can write the momentum
space integral expression that represents the diagram in $D=4-2\epsilon $
dimensions:

\begin{equation}
G=\int \frac{d^{D}q_{1}}{i\pi ^{\frac{D}{2}}}...\frac{d^{D}q_{L}}{i\pi ^{%
\frac{D}{2}}}\frac{1}{(B_{1}^{2}-m_{1}^{2}+i0)^{\nu _{1}}}...\frac{1}{%
(B_{N}^{2}-m_{N}^{2}+i0)^{\nu _{N}}}.  \label{f11}
\end{equation}%
In this expression the $B_{j}$ symbol represents the momentum of the $j$
propagator, whose dependence in general is given as a linear combination of
external and internal momenta. We also define $\left\{ \nu _{1},...,\nu
_{N}\right\} $ as the set of propagator indices or powers, and which are
considered to have arbitrary values. After introducing Schwinger's
parametrization it is possible to evaluate the momentum integrals as
gaussian integrals, and the result of this operation is Schwinger's
parametric representation of $\left( \ref{f11}\right) $, which in the
general case is going to be given by an expression of the form:

\begin{equation}
G=\dfrac{(-1)^{-\frac{LD}{2}}}{\prod\limits_{j=1}^{N}\Gamma (\nu _{j})}%
\dint\limits_{0}^{\infty }d\overrightarrow{x}\;\dfrac{\exp \left(
\sum\limits_{j=1}^{N}x_{j}m_{j}^{2}\right) \exp \left( -\dfrac{F}{U}\right)
}{U^{\frac{D}{2}}}.  \label{f12}
\end{equation}%
For simplicity we have introduced the notation $d\overrightarrow{x}%
=dx_{1}...dx_{N}\;\prod\nolimits_{j=1}^{N}x_{j}^{\nu _{j}-1}$, where $F$ is
defined as:

\begin{equation}
F=\dsum\limits_{i,j=1}^{E}C_{ij}\;p_{i}.p_{j}.
\end{equation}%
The function $U$ and the coefficients $C_{ij}$ are $L$-$lineal$ and $(L+1)$-$%
lineal$ homogeneous polynomials (Symanzik or Kirchhoff polynomials \cite{ViR}
) respectively , in the Schwinger parameters. Both $U$ and $C_{ij}$
correspond to determinants related to a matrix that we have called matrix of
parameters $\mathbf{M}$ \cite{IGo}. An important characteristic of the
coefficients $C_{ij}$ is that they are symmetric $\left(
C_{ij}=C_{ji}\right) $, which is due to the commutativity of the internal
product between the independent external momenta associated to the graph $G$%
. These determinants are defined through the following expressions:

\begin{equation}
\begin{array}{ll}
U= & \left\vert
\begin{array}{ccc}
M_{11} & \cdots & M_{1L} \\
\vdots &  & \vdots \\
M_{L1} & \cdots & M_{LL}%
\end{array}%
\right\vert ,%
\end{array}%
\end{equation}

\begin{equation}
C_{ij}=\left\vert
\begin{tabular}{lllc}
$M_{11}$ & $\cdots $ & $M_{1L}$ & $M_{1(L+j)}$ \\
$\vdots $ &  & $\vdots $ & $\vdots $ \\
\multicolumn{1}{c}{$M_{L1}$} & \multicolumn{1}{c}{$\cdots $} &
\multicolumn{1}{c}{$M_{LL}$} & $M_{L(L+j)}$ \\
$M_{(L+i)1}$ & $\cdots $ & $M_{(L+i)L}$ & $M_{(L+i)(L+j)}$%
\end{tabular}%
\right\vert .
\end{equation}%
As can be seen, the determinants are associated to submatrices of the matrix
of parameters $\mathbf{M}$, which is symmetric and of dimension $\left(
L+E\right) \times \left( L+E\right) $. The actual form of this matrix can be
easily obtained when $\left( \ref{f11}\right) $ is parameterized and the
internal products of all the (internal and external) momenta associated to $%
G $ are expanded, arriving at a quadratic form. The coefficients of such an
expansion correspond to the matrix elements $M_{ij}$. For a better
understanding of this process let us define for convenience the momentum:%
\begin{equation}
Q_{j}=\left\{
\begin{array}{lll}
q_{j} & \text{if} & L\geq j\geq 1 \\
&  &  \\
p_{j-L} & \text{if} & \left( L+E\right) \geq j>L,%
\end{array}%
\right.
\end{equation}%
with which one can build the $\left( L+E\right) $-$vector$ $\mathbf{Q=[}%
Q_{1}\;\;Q_{2}\;...\;Q_{(L+E)}]^{\mathbf{t}}$. Using this definition the
following matrix structure is generated in the integral, after the
parametrization application and before the loop momenta integration:

\begin{equation}
G=\dfrac{1}{\prod\limits_{j=1}^{N}\Gamma (\nu _{j})}\dint\limits_{0}^{\infty
}d\overrightarrow{x}\;\exp \left( \sum\limits_{j=1}^{N}x_{j}m_{j}^{2}\right)
\dint \prod\limits_{j=1}^{L}\left( \frac{d^{D}Q_{j}}{i\pi ^{D/2}}\right)
\exp \left(
-\sum\limits_{i=1}^{L+E}\sum\limits_{j=1}^{L+E}Q_{i}M_{ij}Q_{j}\right) ,
\end{equation}%
and from which we can identify the symmetric matrix $\mathbf{M}$.

\subsection{Foundations of the integration method IBFE}

\subsubsection{Notation and fundamental formulae}

This technique can be introduced directly by considering the integral
expression for the Gamma function. The idea is to obtain operational rules,
which will allow us later on to work with generalized complicated structures
of this type of integrals, such as Schwinger's parametric representation of
a generic Feynman diagram $\left( \ref{f12}\right) $. Let us analyze the
following integral structure:

\begin{equation}
\frac{1}{A^{\beta }}=\frac{1}{\Gamma (\beta )}\dint\limits_{0}^{\infty
}dx\;x^{\beta -1}\exp (-Ax),  \label{f20}
\end{equation}%
where the quantities $A$ and $\beta $ are arbitrary. Expanding the integrand
we get:

\begin{equation}
\frac{1}{A^{\beta }}=\frac{1}{\Gamma (\beta )}\dsum\limits_{n}\phi
_{n}\;A^{n}\dint\limits_{0}^{\infty }dx\;x^{\beta +n-1},  \label{f29}
\end{equation}%
where we have defined the factor:

\begin{equation}
\phi _{n}=\frac{\left( -1\right) ^{n}}{\Gamma \left( n+1\right) }.
\label{f30}
\end{equation}%
Now the integral evaluation will not be done in the usual way, but we define
the following operational relation:

\begin{equation}
\dint\limits_{0}^{\infty }dx\;x^{\beta +n-1}\equiv \Gamma \left( \beta
\right) \dfrac{\Gamma \left( n+1\right) }{\left( -1\right) ^{n}}\;\delta
_{\beta +n,0},  \label{f28}
\end{equation}%
which makes $\left( \ref{f29}\right) $ to be an identity:

\begin{equation}
\begin{array}{ll}
\dfrac{1}{A^{\beta }} & =\dfrac{1}{\Gamma (\beta )}\dsum\limits_{n}\dfrac{%
\left( -1\right) ^{n}}{\Gamma \left( n+1\right) }A^{n}\dint\limits_{0}^{%
\infty }dx\;x^{\beta +n-1}=\dfrac{1}{\Gamma (\beta )}\dsum\limits_{n}\dfrac{%
\left( -1\right) ^{n}}{\Gamma \left( n+1\right) }A^{n}\left( \Gamma \left(
\beta \right) \dfrac{\Gamma \left( n+1\right) }{\left( -1\right) ^{n}}%
\;\delta _{\beta +n,0}\right) \\
&  \\
& =\dsum\limits_{n}A^{n}\;\delta _{\beta +n,0} \\
&  \\
& =A^{-\beta }.%
\end{array}%
\end{equation}%
For convenience we introduce the following notation:

\begin{equation}
\dint dx\;x^{\left( \alpha +\beta \right) -1}\equiv \left\langle \alpha
+\beta \right\rangle ,  \label{f21}
\end{equation}%
where the parenthesis $\left\langle \cdot \right\rangle $ has implicit the
constraint associated to the Kronecker delta. This identity is the first
fundamental formula of the integration method IBFE. With the previously
defined notation, equation $\left( \ref{f29}\right) $ can be written in the
following way:

\begin{equation}
\frac{1}{A^{\beta }}=\frac{1}{\Gamma (\beta )}\dsum\limits_{n}\phi
_{n}\;A^{n}\left\langle \beta +n\right\rangle ,
\end{equation}%
which corresponds to the multiregion expansion (MRE) of the factor $%
A^{-\beta }$.

On the other hand, starting from the application of equation $\left( \ref%
{f20}\right) $ and later equation $\left( \ref{f21}\right) $ to an arbitrary
multinomial, we find the second fundamental formula that the method uses.
This expresses the fact that a multinomial of $\sigma $ terms can be written
as a MRE, in such a way that it contains simultaneously all the possible
expansions with respect to the ratio of the different terms that are present
in the multinomial. This MRE can be written in the following form:

\begin{equation}
\left( A_{1}+...+A_{\sigma }\right) ^{\pm \nu
}=\dsum\limits_{n_{1}}...\dsum\limits_{n_{\sigma }}\phi _{n_{1},..,n_{\sigma
}}\;A_{1}^{n_{1}}...A_{\sigma }^{n_{\sigma }}\frac{\left\langle \mp \nu
+n_{1}+...+n_{\sigma }\right\rangle }{\Gamma (\mp \nu )},  \label{f22}
\end{equation}%
where the definition of the factor $\left( \ref{f30}\right) $ has been
generalized to:%
\begin{equation}
\phi _{n_{1},..,n_{\sigma }}=\phi _{n_{1}}...\phi _{n_{\sigma
}}=(-1)^{_{n_{1}+...+n_{\sigma }}}\dfrac{1}{\Gamma (n_{1}+1)...\Gamma
(n_{\sigma }+1)}.
\end{equation}

\subsubsection{General form of a diagram MRE and its solutions}

Once the parametric representation of the diagram $\left( \ref{f12}\right) $
has been obtained, the following step is finding its MRE, and for this it is
necessary to expand the integrand starting with the exponentials if they
exist, and then expanding all the multinomials that the procedure is
generating according to formula $\left( \ref{f22}\right) $. This expansion
process stops when finally there is only one term, which is a product of all
the Schwinger parameters. At this point all that is left is to replace the
integrals according to formula $\left( \ref{f21}\right) $\ into its
equivalent $\left\langle \cdot \right\rangle $. The result is the MRE of the
Feynman integral $G$ considered in $\left( \ref{f11}\right) $. In the case
of a general topology characterized by $\mathit{M}$ different mass scales, $%
\mathit{N}$ propagators, $L$ loops and a minimal quantity $\mathit{P}$ of
invariants associated to scalar products of the independent external
momenta, then the general form of the MRE of $G$ is given by the following
expression:

\begin{equation}
G=(-1)^{-\frac{LD}{2}}\dsum\limits_{n_{1},..,n_{\sigma }}\phi
_{n_{1},..,n_{\sigma }}\;\tprod\limits_{i=1}^{\mathit{P}}(Q_{i}^{2})^{n_{i}}%
\tprod\limits_{j=P+1}^{\mathit{P+M}}(-m_{j}^{2})^{n_{j}}\tprod\limits_{k=1}^{%
\mathit{N}}\dfrac{\left\langle \nu _{k}+\alpha _{k}\right\rangle }{\Gamma
(\nu _{k})}\tprod\limits_{l=1}^{\mathit{K}}\dfrac{\left\langle \beta
_{l}+\gamma _{l}\right\rangle }{\Gamma (\beta _{l})},  \label{f23}
\end{equation}%
where we can identify the following quantities:

\begin{itemize}
\item $\sigma \Longrightarrow $ Multiplicity or number of sums that are
present in the MRE of diagram $G$.

\item $Q_{j}^{2}\Longrightarrow $ A kinematical invariant, which is a
quadratic form of the independent external momenta.

\item $\alpha _{j},\beta _{j},\gamma _{j}\Longrightarrow $ Linear
combinations of the indexes $\left\{ n_{1},..,n_{\sigma }\right\} $, except
for $\beta _{1}$, which is the only one that has dependence on the dimension
$D$:

\begin{equation}
\beta _{1}=\frac{D}{2}+n_{1}+...+n_{P}.
\end{equation}%
The coefficients of the indices of the sum $\left\{ n_{i}\right\} $ in the
linear combinations $\alpha _{j}$ and $\gamma _{j}$ are $(+1)$, and in the
case of $\beta _{j}$ the indices have coefficients $(-1)$, except for $\beta
_{1}$.

\item $\mathit{N}\Longrightarrow $ Number of propagators or equivalently
number of parametric integrations, which the method transforms into $\mathit{%
N}$ Kronecker deltas.

\item $\mathit{K}\Longrightarrow $ Total number of MREs performed over the
integrand of the parametric representation, which in turn generates $\mathit{%
K}$ constraints or equivalently $\mathit{K}$ Kronecker deltas. This only
refers to multinomial expansions that are present in the parametric integral.
\end{itemize}

\bigskip

In order to find the solutions it is necessary to evaluate the sums that are
present in $\left( \ref{f23}\right) $, using for this purpose the existing
constraints between the sum indices, represented by the $\delta =(\mathit{N+K%
})$ Kronecker deltas. There are several ways to do this evaluation, and in
fact the number of different ways to evaluate the MRE of $G$ using the
Kronecker deltas is given by the combinatorial formula:

\begin{equation}
C_{\delta }^{\sigma }=\dfrac{\sigma !}{\delta !(\sigma -\delta )!}.
\end{equation}%
Each of these forms of summing will generate as a result a term in the
solution of $G$, which corresponds to a generalized hypergeometric function,
and whose multiplicity is given by:

\begin{equation}
\mu =\left( \sigma -\delta \right) .
\end{equation}%
In general it is not always possible to use the $\delta $\ Kronecker deltas
to evaluate a similar number of sums, since this will depend on the
combination of sum indices for which the sum is going to be done. If this
happens, these cases simply do not generate contributions to the solution.

In simple terms, the idea of the method is to generate finally an expansion
that represents the diagram $G$, the MRE of $G$, characterized by a
multiplicity $\sigma $ in combination with $\delta$ Kronecker deltas. From
this it is possible to get the solution in terms of generalized
hypergeometric functions, series of multiplicity $\mu$ whose arguments
usually correspond to ratios between two characteristic energy scales of the
diagram, or in the more general case include sums of unitary argument.

\section{Conventional reduction of diagrams with massless bubble insertions}

\subsection{Graphical representation of the reduction}

Many diagrams contain bubble type insertions in the propagators, which in
the massless case can be directly reduced to a propagator and a factor which
is a ratio of Gamma functions. This result is straightforward and is very
useful for an iterative or loop by loop reduction of topologies or
subtopologies built in terms of bubble type insertions. The fundamental
formula for insertions or bubble modules can be deduced directly from the
solution of the momentum integral for this diagram:

\begin{equation}
G=\dint \frac{d^{D}q}{i\pi ^{D/2}}\frac{1}{(q^{2})^{a_{1}}\left(
(q+p)^{2}\right) ^{a_{2}}},  \label{ff1}
\end{equation}%
where the indices $a_{1}$ and $a_{2}$ are arbitrary. The explicit solution
of this integral is:

\begin{equation}
G=g(a_{1},a_{2})\;\dfrac{1}{(p^{2})^{a_{1}+a_{2}-\frac{D}{2}}},  \label{ff2}
\end{equation}%
where the factor $g(a_{1},a_{2})$ is given by:

\begin{equation}
g(a_{1},a_{2})=(-1)^{-\frac{D}{2}}\dfrac{\Gamma (a_{1}+a_{2}-\frac{D}{2}%
)\Gamma (\frac{D}{2}-a_{1})\Gamma (\frac{D}{2}-a_{2})}{\Gamma (a_{1})\Gamma
(a_{2})\Gamma (D-a_{1}-a_{2})}.  \label{ff17}
\end{equation}%
Usually equation $\left( \ref{ff2}\right) $ is represented pictorially in
the following way:

\begin{equation}
\begin{array}{cc}
\quad a_{1} &  \\
\begin{minipage}{4.4cm} \includegraphics[scale=.7]
{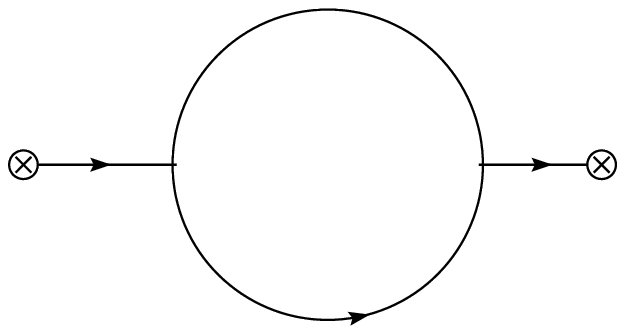} \end{minipage} &
\begin{array}{cc}
& a_{1}+a_{2}-\frac{D}{2} \\
=\;g(a_{1},a_{2})\;\times &
\begin{minipage}{3.3cm} \includegraphics[scale=.8]
{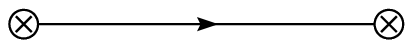} \end{minipage} \\
&
\end{array}
\\
\quad a_{2} &
\end{array}
\label{ff11}
\end{equation}%
Another graphical formula, which is useful when two or more scalar
propagators are in series, is:

\begin{equation}
\begin{array}{ccc}
a_{1}\qquad a_{2} &  & \left( a_{1}+a_{2}\right) \\
\begin{minipage}{3.3cm} \includegraphics[scale=.8]
{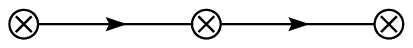} \end{minipage} & = &
\begin{minipage}{3.3cm} \includegraphics[scale=.8]
{graph_D.eps} \end{minipage}%
\end{array}
\label{ff4}
\end{equation}%
This last formula is also valid for a massive theory in the case of two
equal masses. This two pictorial expressions are enough in order to find the
solution of this family of diagrams. Finally the operational problem
consists in literally reducing series and parallel propagators, and the
formulae $\left( \ref{ff11}\right) $ and $\left( \ref{ff4}\right) $
represent this operation.

\subsection{A simple application}

In order to show the usefulness of the previous pictorial formulae $\left( %
\ref{ff11}\right) $\ and $\left( \ref{ff4}\right) $, let us consider the
following radiative correction to a propagator through which it flows a
momentum $p$:

\begin{equation}
G=%
\begin{minipage}{4.7cm} \includegraphics[scale=.7]
{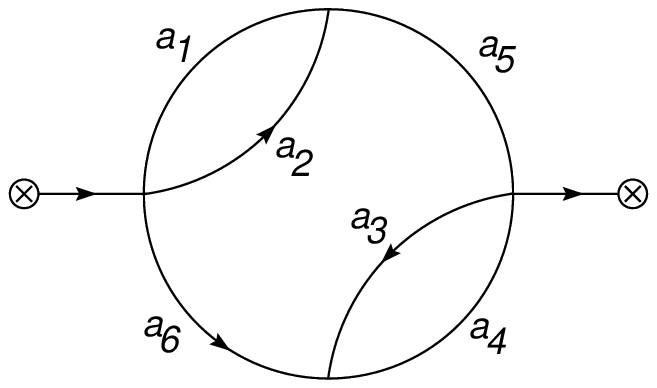} \end{minipage}
\end{equation}%
Then, using equation $\left( \ref{ff11}\right) $ we can reduce the two
bubble type subgraphs that appear in the diagram, and we get:

\begin{equation}
G=g(a_{1},a_{2})\times g(a_{3},a_{4})\;\times
\begin{minipage}{5.2cm} \includegraphics[scale=.7]
{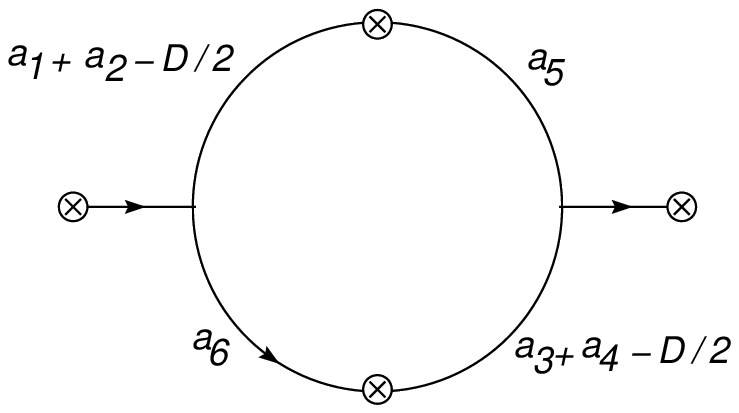} \end{minipage}
\end{equation}%
The resulting loop is easily evaluated after using equation $\left( \ref{ff4}%
\right) $, and this finally allows us to get the solution of this diagram:

\begin{equation}
\begin{array}{cc}
& a_{1}+...+a_{6}-\frac{3D}{2} \\
G=g(a_{1},a_{2})\times g(a_{3},a_{4})\times g(a_{1}+a_{2}+a_{5}-\tfrac{D}{2}%
,a_{3}+a_{4}+a_{6}-\tfrac{D}{2})\;\times &
\begin{minipage}{3.3cm} \includegraphics[scale=.8]
{graph_D.eps} \end{minipage},%
\end{array}%
\end{equation}%
which is equivalent to the following expression:

\begin{equation}
\fbox{$G=g(a_{1},a_{2})\times g(a_{3},a_{4})\times g(a_{125}-\tfrac{D}{2}%
,a_{346}-\tfrac{D}{2})\;\dfrac{1}{(p^{2})^{a_{1}+...+a_{6}-\frac{3D}{2}}}$}
\label{ff7}
\end{equation}%
To simplify the notation we have defined for the sum of indices the
following symbology:

\begin{equation}
a_{ijk...}=a_{i}+a_{j}+a_{k}+...
\end{equation}%
Using then formula $\left( \ref{ff17}\right) $ we finally obtain the
explicit solution for the diagram:

\begin{equation}
\begin{array}{ll}
G= & (-1)^{-\frac{3D}{2}}\dfrac{\Gamma (a_{12}-\frac{D}{2})\Gamma (\frac{D}{2%
}-a_{1})\Gamma (\frac{D}{2}-a_{2})}{\Gamma (a_{1})\Gamma (a_{2})\Gamma
(D-a_{12})}\dfrac{\Gamma (a_{34}-\frac{D}{4})\Gamma (\frac{D}{4}%
-a_{3})\Gamma (\frac{D}{4}-a_{4})}{\Gamma (a_{3})\Gamma (a_{4})\Gamma
(D-a_{34})} \\
&  \\
& \dfrac{\Gamma (a_{123456}-\frac{3D}{2})\Gamma (D-a_{125})\Gamma (D-a_{346})%
}{\Gamma (a_{125}-\tfrac{D}{2})\Gamma (a_{346}-\tfrac{D}{2})\Gamma
(2D-a_{123456})}\dfrac{1}{(p^{2})^{a_{1}+...+a_{6}-\frac{3D}{2}}}.%
\end{array}%
\end{equation}

\section{IBFE and modular reduction of diagrams}

\subsection{The bubble module}

The application of IBFE to the bubble module will allow us to find the MREs
that generate the $1$-$loop$ functions which identify the different masses
and momenta configurations that can be associated to this diagram. We start
our analysis by considering this diagram to be associated to the kinematical
variable $p^{2}$, and then from this we can define the following associated $%
1$-$loop$ functions: $G_{k}$ and $\overline{G}_{k}\;\left( k=A,B,C,D\right) $%
, according to the value of this variable and the corresponding mass
configuration for this module:

\begin{itemize}
\item Case $p^{2}\neq 0$

$G_{A}\implies $ Massless propagators.

$G_{B}\implies $ One massive propagator.

$G_{C}\implies $ Equal mass propagators.

$G_{D}\implies $ Different mass propagators.

\item Case $p^{2}=0$ : This could be obtained from the previous functions,
just by taking the kinematical variable to zero. Nevertheless, we will take
a different approach, which uses the series multiregion representation
obtained directly from the parametric representation of a vacuum bubble. The
reason is simple, in general the mathematical expression for the MRE is
simpler than the corresponding expression obtained by taking $p^{2}=0$ in
the functions $G_{k}$. These functions are useful when diagrams associated
with vacuum fluctuations are evaluated. The $1$-$loop$ functions that we
will define for this case are analogous to the previous ones:\bigskip

$\overline{G}_{A}\implies $ Massless propagators.

$\overline{G}_{B}\implies $ One massive propagator.

$\overline{G}_{C}\implies $ Equal mass propagators.

$\overline{G}_{D}\implies $ Different mass propagators.

\bigskip
\end{itemize}

For our analysis let us start by writing the general integral representation
of this diagram in momentum space:

\begin{equation}
G=\dint \frac{d^{D}q}{i\pi ^{D/2}}\frac{1}{(q^{2}-M_{j}^{2})^{a_{j}}\left(
(p-q)^{2}-M_{k}^{2}\right) ^{a_{k}}}.
\end{equation}%
We shall consider general cases, so the powers $a_{j}$ and $a_{k}$ are
arbitrary. We then obtain Schwinger's parametric representation:

\begin{equation}
G=\dfrac{(-1)^{-\frac{D}{2}}}{\Gamma (a_{j})\Gamma (a_{k})}%
\dint\limits_{0}^{\infty }d\overrightarrow{x}\;\frac{\exp
(x_{j}M_{j}^{2})\exp \left( x_{k}M_{k}^{2}\right) \exp \left( -\dfrac{%
x_{j}x_{k}}{x_{j}+x_{k}}p^{2}\right) }{\left( x_{j}+x_{k}\right) ^{\frac{D}{2%
}}}.  \label{ff21}
\end{equation}%
Depending on the specific values of the masses $\left\{ M_{j},M_{k}\right\} $
and of $\left\{ p^{2}\right\} $, we can consider eight possible $1$-$loop$
functions, which will be discussed in what follows.

\subsection{$1$-$loop$ functions for $p^{2}\neq 0$}

There are four loop functions that can be defined in this case ($%
G_{A},G_{B},G_{C}$ and $G_{D}$), of which only the last two are independent
. The loop function $G_{A}$ is a particular case of $G_{C}$ or $G_{D}$, and
in the same way $G_{B}$ turns out to be a particular case of $G_{D}$. This
will be shown now.

\subsubsection{$1$-$loop$ function $G_{D}\Longrightarrow M_{j}\neq M_{k}\neq
0$}

Let us start with the most general case, with propagators with different
masses. Then the corresponding parametric integral is given through equation
$\left( \ref{ff21}\right) $, and the first step is to fractionally expand
the exponential that contains the invariant $p^{2}$, which gives:

\begin{equation}
G=\dfrac{(-1)^{-\frac{D}{2}}}{\Gamma (a_{j})\Gamma (a_{k})}%
\dsum\limits_{n}\phi _{n}\;\left( p^{2}\right) ^{n}\dint\limits_{0}^{\infty
}d\overrightarrow{x}\;\exp (x_{j}M_{j}^{2})\exp \left( x_{k}M_{k}^{2}\right)
\frac{x_{j}^{n}x_{k}^{n}}{\left( x_{j}+x_{k}\right) ^{\frac{D}{2}+n}}.
\end{equation}%
or equivalently:

\begin{equation}
G=\dsum\limits_{n}G_{D}(a_{j},a_{k};n;M_{j}^{2},M_{k}^{2})\;\left(
p^{2}\right) ^{n}=\dsum\limits_{n}G_{D}(a_{j},a_{k};n;M_{j}^{2},M_{k}^{2})\;%
\frac{1}{\left( p^{2}\right) ^{-n}},  \label{ff29}
\end{equation}%
where the $1$-$loop$ function has been defined as:

\begin{equation}
G_{D}(a_{j},a_{k};n;M_{j}^{2},M_{k}^{2})=\dfrac{(-1)^{-\frac{D}{2}}}{\Gamma
(a_{j})\Gamma (a_{k})}\phi _{n}\;\dint\limits_{0}^{\infty }d\overrightarrow{x%
}\;\exp (x_{j}M_{j}^{2})\exp \left( x_{k}M_{k}^{2}\right) \frac{%
x_{j}^{n}x_{k}^{n}}{\left( x_{j}+x_{k}\right) ^{\frac{D}{2}+n}}.
\label{ff23}
\end{equation}%
Remember that the factor $\phi _{n}$ is given by $\phi _{n}=\dfrac{%
(-1)^{_{n}}}{\Gamma (n+1)}$. The following step is finding the MRE of the
loop function $G_{D}(a_{j},a_{k};n;M_{j}^{2},M_{k}^{2})$, and for this
purpose we expand the mass exponentials in $\left( \ref{ff23}\right) $ and
then the binomial in the denominator of the same formula, which gives:

\begin{equation}
\dfrac{1}{\left( x_{j}+x_{k}\right) ^{\frac{D}{2}+n}}=\dsum%
\limits_{l_{j},l_{k}}\phi _{l_{j},l_{k}}\;x_{j}^{l_{j}}x_{k}^{l_{k}}\frac{%
\left\langle \tfrac{D}{2}+n+l_{j}+l_{k}\right\rangle }{\Gamma (\frac{D}{2}+n)%
},  \label{ff12}
\end{equation}%
Performing the necessary algebra, the MRE for $%
G_{D}(a_{j},a_{k};n;M_{j}^{2},M_{k}^{2})$ is finally given by:

\begin{equation}
\fbox{$G_{D}(a_{j},a_{k};n;M_{j}^{2},M_{k}^{2})=\dfrac{(-1)^{-\frac{D}{2}}}{%
\Gamma (a_{j})\Gamma (a_{k})}\dsum\limits_{\substack{ m_{j},m_{k}  \\ %
l_{j},l_{k}}}\phi _{n,m_{j},m_{k},l_{j},l_{k}}\;(-M_{j}^{2})^{m_{j}}\left(
-M_{k}^{2}\right) ^{m_{k}}\dfrac{\Delta _{1}\Delta _{2}\Delta _{3}}{\Gamma (%
\frac{D}{2}+n)}$}  \label{ff24}
\end{equation}%
with the constraints being defined by the identities:

\begin{equation}
\left\{
\begin{array}{l}
\Delta _{1}=\left\langle \frac{D}{2}+n+l_{j}+l_{k}\right\rangle , \\
\Delta _{2}=\left\langle a_{j}+n+m_{j}+l_{j}\right\rangle , \\
\Delta _{3}=\left\langle a_{k}+n+m_{k}+l_{k}\right\rangle .%
\end{array}%
\right.
\end{equation}%
The process of replacing the graph in terms of this equivalent $1$-$loop$
function (see equation $\left( \ref{ff29}\right) $ ) can be written
symbolically through the following graphical formulation:

\begin{equation}
\fbox{$%
\begin{array}{cc}
\;a_{j},M_{j} &  \\
\begin{minipage}{4.5cm} \includegraphics[scale=.7]
{graph_A.eps} \end{minipage} & =%
\begin{array}{cc}
& -n \\
\dsum\limits_{n}G_{D}(a_{j},a_{k};n;M_{j}^{2},M_{k}^{2})\;\times &
\begin{minipage}{3.3cm} \includegraphics[scale=.8]
{graph_D.eps} \end{minipage} \\
&
\end{array}
\\
\;a_{k},M_{k} &
\end{array}%
$}
\end{equation}

\paragraph{Derivation of $G_{A}$ from $G_{D}$ $\left( \text{Case }%
M_{j}=M_{k}=0\right) $ :}

The loop function $G_{A}$ turns out to be a particular case of the $1$-$loop$
function $G_{D}$. The function $G_{A}$ can be defined as:

\begin{equation}
G_{A}\left( a_{j},a_{k};n\right) =G_{D}(a_{j},a_{k};n;0,0),
\end{equation}%
and whose equivalent MRE corresponds to the expression:

\begin{equation}
\fbox{$G_{A}\left( a_{j},a_{k};n\right) =\dfrac{(-1)^{-\frac{D}{2}}}{\Gamma
(a_{j})\Gamma (a_{k})}\dsum\limits_{l_{j},l_{k}}\phi _{n,l_{j},l_{k}}\;%
\dfrac{\Delta _{1}\Delta _{2}\Delta _{3}}{\Gamma (\frac{D}{2}+n)}$}
\label{ff5}
\end{equation}%
where the masses were put to zero in $\left( \ref{ff24}\right) $ and the
sums associated to the indices $\left\{ m_{j},m_{k}\right\} $ were
eliminated, taking later on these indices to zero. The constraints are now
given by the identities:

\begin{equation}
\left\{
\begin{array}{l}
\Delta _{1}=\left\langle \frac{D}{2}+n+l_{j}+l_{k}\right\rangle , \\
\Delta _{2}=\left\langle a_{j}+n+l_{j}\right\rangle , \\
\Delta _{3}=\left\langle a_{k}+n+l_{k}\right\rangle .%
\end{array}%
\right.
\end{equation}%
As before it is possible to represent the fractional expansion of the graph
symbolically through the following graphical equation:

\begin{equation}
\fbox{$%
\begin{array}{cc}
a_{j} &  \\
\begin{minipage}{4.5cm} \includegraphics[scale=.7]
{graph_A.eps} \end{minipage} & =%
\begin{array}{cc}
& -n \\
\dsum\limits_{n}G_{A}\left( a_{j},a_{k};n\right) \;\times &
\begin{minipage}{3.3cm} \includegraphics[scale=.8]
{graph_D.eps} \end{minipage} \\
&
\end{array}
\\
a_{k} &
\end{array}%
$}  \label{ff3}
\end{equation}

\paragraph{Derivation of $G_{B}$ $\left( \text{Case }M_{j}\neq
0,M_{k}=0\right) $ :}

The loop function $G_{B}$ can be written in terms of the loop function $%
G_{D} $ as:

\begin{equation}
G_{B}(a_{j},a_{k};n;M_{j}^{2})=G_{D}(a_{j},a_{k};n;M_{j}^{2},0),
\end{equation}%
the corresponding MRE is obtained just as for $G_{A}$:

\begin{equation}
\fbox{$G_{B}(a_{j},a_{k};n;M_{j}^{2})=\dfrac{(-1)^{-\frac{D}{2}}}{\Gamma
(a_{j})\Gamma (a_{k})}\dsum\limits_{m_{j},l_{j},l_{k}}\phi
_{n,m_{j},l_{j},l_{k}}\;(-M_{j}^{2})^{m_{j}}\dfrac{\Delta _{1}\Delta
_{2}\Delta _{3}}{\Gamma (\frac{D}{2}+n)}$}
\end{equation}%
The constraints now are given by:

\begin{equation}
\left\{
\begin{array}{l}
\Delta _{1}=\left\langle \frac{D}{2}+n+l_{j}+l_{k}\right\rangle , \\
\Delta _{2}=\left\langle a_{j}+n+m_{j}+l_{j}\right\rangle , \\
\Delta _{3}=\left\langle a_{k}+n+l_{k}\right\rangle .%
\end{array}%
\right.
\end{equation}%
The graphical equation which represents the loop reduction is now:

\begin{equation}
\fbox{$%
\begin{array}{cc}
a_{j},M_{j} &  \\
\begin{minipage}{4.5cm} \includegraphics[scale=.7]
{graph_A.eps} \end{minipage} & =%
\begin{array}{cc}
& -n \\
\dsum\limits_{n}G_{B}(a_{j},a_{k};n;M_{j}^{2})\;\times &
\begin{minipage}{3.3cm} \includegraphics[scale=.8]
{graph_D.eps} \end{minipage} \\
&
\end{array}
\\
a_{k} &
\end{array}%
$}  \label{ff9}
\end{equation}

\subsubsection{$1$-$loop$ function $G_{C}\Longrightarrow M_{j}=M_{k}=M$}

The loop function $G_{C}$ does not correspond to a particularization of $%
G_{D}$. To see this we start with the parametric representation already
factorized:

\begin{equation}
G=\dfrac{(-1)^{-\frac{D}{2}}}{\Gamma (a_{j})\Gamma (a_{k})}%
\dint\limits_{0}^{\infty }d\overrightarrow{x}\;\frac{\exp \left(
UM^{2}\right) \exp \left( -\dfrac{x_{j}x_{k}}{U}p^{2}\right) }{U^{\frac{D}{2}%
}},  \label{ff25}
\end{equation}%
with $U=\left( x_{j}+x_{k}\right) $. The reason that this loop function
cannot be derived from $G_{D}$ is related to the denominator structure,
where the factorization of the polynomial $U$ produces an MRE of $\left( \ref%
{ff25}\right) $ which is more reduced than when we take $M_{j}=M_{k}$ in $%
G_{D}$.

In equation $\left( \ref{ff25}\right) $ we expand the exponential which
contains the momentum, and obtain the following series:

\begin{equation}
G=\dsum\limits_{n}G_{C}\left( a_{j},a_{k};n;M^{2}\right) \;\frac{1}{\left(
p^{2}\right) ^{-n}},
\end{equation}%
where we have defined the $1$-$loop$ function $G_{C}\left(
a_{j},a_{k};n;M^{2}\right) $:

\begin{equation}
G_{C}\left( a_{j},a_{k};n;M^{2}\right) =\dfrac{(-1)^{-\frac{D}{2}}}{\Gamma
(a_{j})\Gamma (a_{k})}\phi _{n}\;\dint\limits_{0}^{\infty }d\overrightarrow{x%
}\;\exp \left( \left( x_{j}+x_{k}\right) M^{2}\right) \frac{%
x_{j}^{n}x_{k}^{n}}{\left( x_{j}+x_{k}\right) ^{\frac{D}{2}+n}},
\end{equation}%
and expanding the exponential which contains the mass term we have the
series:

\begin{equation}
G_{C}\left( a_{j},a_{k};n;M^{2}\right) =\dfrac{(-1)^{-\frac{D}{2}}}{\Gamma
(a_{j})\Gamma (a_{k})}\phi _{n}\;\dsum\limits_{l_{i}}\phi _{l_{i}}\;\left(
-M^{2}\right) ^{l_{i}}\dint\limits_{0}^{\infty }d\overrightarrow{x}\;\frac{%
x_{j}^{n}x_{k}^{n}}{\left( x_{j}+x_{k}\right) ^{\frac{D}{2}+n-l_{i}}}.
\label{ff13}
\end{equation}%
Now we shall find the MRE for the binomial in the integrand denominator:

\begin{equation}
\dfrac{1}{\left( x_{j}+x_{k}\right) ^{\frac{D}{2}+n-l_{i}}}%
=\dsum\limits_{l_{j},l_{k}}\phi _{l_{j},l_{k}}\;x_{j}^{l_{j}}x_{k}^{l_{k}}%
\frac{\left\langle \frac{D}{2}+n-l_{i}+l_{j}+l_{k}\right\rangle }{\Gamma (%
\frac{D}{2}+n-l_{i})},
\end{equation}%
which replaced in $\left( \ref{ff13}\right) $ and after some algebra allows
finally to get the required MRE:

\begin{equation}
\fbox{$G_{C}\left( a_{j},a_{k};n;M^{2}\right) =\dfrac{(-1)^{-\frac{D}{2}}}{%
\Gamma (a_{j})\Gamma (a_{k})}\dsum\limits_{l_{i},l_{j},l_{k}}\phi
_{n,l_{i},l_{j},l_{k}}\;(-M_{j}^{2})^{l_{i}}\dfrac{\Delta _{1}\Delta
_{2}\Delta _{3}}{\Gamma (\frac{D}{2}+n-l_{i})}$}
\end{equation}%
where the constraints $\left\{ \Delta _{k}\right\} $\ are given by:

\begin{equation}
\left\{
\begin{array}{l}
\Delta _{1}=\left\langle \frac{D}{2}+n-l_{i}+l_{j}+l_{k}\right\rangle , \\
\Delta _{2}=\left\langle a_{j}+n+l_{j}\right\rangle , \\
\Delta _{3}=\left\langle a_{k}+n+l_{k}\right\rangle .%
\end{array}%
\right.
\end{equation}%
Then the loop reduction in this case is represented graphically by the
following pictorial equation:

\begin{equation}
\fbox{$%
\begin{array}{cc}
a_{j},M &  \\
\begin{minipage}{4.5cm} \includegraphics[scale=.7]
{graph_A.eps} \end{minipage} & =%
\begin{array}{cc}
& -n \\
\dsum\limits_{n}G_{C}\left( a_{j},a_{k};n;M^{2}\right) \;\times &
\begin{minipage}{3.3cm} \includegraphics[scale=.8]
{graph_D.eps} \end{minipage} \\
&
\end{array}
\\
a_{k},M &
\end{array}%
$}  \label{ff8}
\end{equation}%
Notice that the $1$-$loop$ function $G_{A}$ is also a particular case of $%
G_{C}$:

\begin{equation}
G_{A}\left( a_{j},a_{k};n\right) =G_{C}\left( a_{j},a_{k};n;0\right) .
\end{equation}%
In all the previous representations or reductions of the bubble diagram one
can see the similitude with the conventional equation $\left( \ref{ff11}%
\right) $. Nevertheless, the differences are important since in the
integration by fractional expansion technique (IBFE) massive graphs can be
reduced with the same degree of difficulty as in the massless cases.

\section{MREs for one loop vacuum diagrams}

\qquad The following $1$-$loop$ functions are useful when we wish to
evaluate certain topologies associated with vacuum fluctuations. These
functions will be identified as $\overline{G}_{K}\left( K=A,B,C,D\right) $,
in analogy with the ones obtained in the previous section for $p^{2}\neq 0$.
The corresponding MREs of these new cases will be obtained directly from
their momentum integrals, which corresponds to a bubble without external
lines. The resulting MRE for each $1$-$loop$ function is in fact simpler
than the one obtained by making $p^{2}=0$ in the respective loop function $%
G_{k}$, that is, it has a more reduced composition of sums and Kronecker
deltas.

\subsection{MRE of a bubble with one propagator}

\subsubsection{Massless propagator}

This case does not correspond to any of the bubble diagram formulations
previously obtained because it is a bubble with just one propagator. The
momentum integral is:

\begin{equation}
\begin{array}{cc}
a &  \\
\begin{minipage}{2.3cm} \includegraphics[scale=.7]
{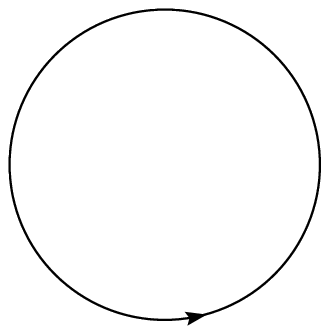} \end{minipage} & =\dint \dfrac{d^{D}q}{i\pi ^{\frac{D}{2}}}%
\dfrac{1}{\left( q^{2}\right) ^{a}}.%
\end{array}
\label{ff30}
\end{equation}%
The representation in terms of MRE is for this case easy to find, using:

\begin{equation}
\begin{array}{ll}
\dint \dfrac{d^{D}q}{i\pi ^{\frac{D}{2}}}\dfrac{1}{\left( q^{2}\right) ^{a}}
& =\dfrac{1}{\Gamma (\nu )}\dint\limits_{0}^{\infty }dx\;x^{a-1}\dint \dfrac{%
d^{D}q}{i\pi ^{\frac{D}{2}}}\;\exp (-xq^{2}) \\
&  \\
& =(-1)^{-\frac{D}{2}}\dfrac{\left\langle a-\frac{D}{2}\right\rangle }{%
\Gamma (a)},%
\end{array}%
\end{equation}%
where we have used the Minskowski space integral identity:

\begin{equation}
\dint \dfrac{d^{D}q}{i\pi ^{\frac{D}{2}}}\;\exp (-\beta q^{2})=\dfrac{(-1)^{-%
\frac{D}{2}}}{\beta ^{\frac{D}{2}}}.
\end{equation}%
Finally the required MRE can be written in terms of a pictorial equation as:

\begin{equation}
\fbox{$%
\begin{array}{c}
a \\
\begin{minipage}{2.3cm} \includegraphics[scale=.7]
{graph_C.eps} \end{minipage} \\
\end{array}%
=(-1)^{-\frac{D}{2}}\dfrac{\left\langle a-\tfrac{D}{2}\right\rangle }{\Gamma
(a)}$}  \label{ff31}
\end{equation}%
One of the properties of Feynman integrals is that they are invariant under
momenta scaling, which allows to show that the integral in equation $\left( %
\ref{ff30}\right) $ vanishes, nevertheless, an MRE has been obtained for
this diagram. The explanation is quite simple. In fact, first we need to
know whether the module constitutes by itself a diagram, or it is s product
of a reduction of a more complex diagram. For the last option the MRE does
not vanish, but has only validity within the integration method that is
used. This allows to formulate a generalization associated to what has been
said previously in terms of a theorem:

\bigskip

\begin{equation*}
\fbox{$%
\begin{array}{l}
\text{Any MRE\textbf{\ }obtained from a Feynman diagram considering all its
loops,} \\
\text{ with a number of constraints\ (Kronecker deltas) } \\
\text{ bigger than the number of summations, vanishes identically.}%
\end{array}%
$}
\end{equation*}

\bigskip

Let us consider a simple example, a two loop vacuum fluctuation in a
massless theory, whose conventional evaluation is rather obvious, but which
nevertheless can be an illustrative example of the theorem:

\begin{equation}
G=%
\begin{minipage}{5cm} \includegraphics[scale=.7]
{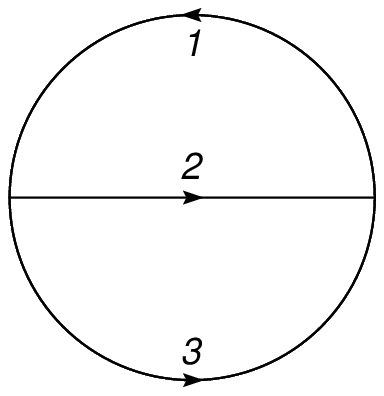} \end{minipage}
\end{equation}%
The parametric representation is given by:

\begin{equation}
G=\dfrac{(-1)^{-D}}{\Gamma (a_{1})\Gamma (a_{2})\Gamma (a_{3})}%
\dint\limits_{0}^{\infty }d\overrightarrow{x}\;\frac{1}{\left(
x_{1}x_{2}+x_{1}x_{3}+x_{2}x_{3}\right) ^{\frac{D}{2}}}.
\end{equation}%
Expanding the integrand denominator:

\begin{equation}
\frac{1}{\left( x_{1}x_{2}+x_{1}x_{3}+x_{2}x_{3}\right) ^{\frac{D}{2}}}%
=\dsum\limits_{n_{1},n_{2},n_{3}}\phi
_{n_{1},n_{2},n_{3}}%
\;x_{1}^{n_{1}+n_{2}}x_{2}^{n_{1}+n_{3}}x_{3}^{n_{2}+n_{3}}\frac{%
\left\langle \tfrac{D}{2}+n_{1}+n_{2}+n_{3}\right\rangle }{\Gamma (\tfrac{D}{%
2})},
\end{equation}%
and replacing the integral signs by their respective constraints we obtain
the following associated MRE:

\begin{equation}
G=\dfrac{(-1)^{-D}}{\Gamma (a_{1})\Gamma (a_{2})\Gamma (a_{3})}%
\dsum\limits_{n_{1},n_{2},n_{3}}\phi _{n_{1},n_{2},n_{3}}\;\frac{%
\left\langle \tfrac{D}{2}+n_{1}+n_{2}+n_{3}\right\rangle \left\langle
a_{1}+n_{1}+n_{2}\right\rangle \left\langle a_{2}+n_{1}+n_{3}\right\rangle
\left\langle a_{3}+n_{2}+n_{3}\right\rangle }{\Gamma (\tfrac{D}{2})}.
\end{equation}%
The number of constraints (Kronecker deltas) is bigger than the number of
summations, and therefore the previously mentioned theorem says that it
should vanish:

\begin{equation}
G\equiv 0.
\end{equation}

\subsubsection{Massive propagator}

The next degree of difficulty comes from the addition of a mass to the
propagator, for which we will use two different ways of finding the
corresponding MRE. In the first one the usual formalism will be employed,
which consists in parametrizing and then replacing the resulting integral
for its corresponding MRE. The second relies in expressing the propagator in
terms of its own MRE, and then parametrize the loop integral which now does
not contain any mass term.

\paragraph{Alternative I : Direct parametrization of the integral}

We want to find the MRE of the following Feynman integral:

\begin{equation}
\begin{array}{cc}
a,m &  \\
\begin{minipage}{2.3cm} \includegraphics[scale=.7]
{graph_C.eps} \end{minipage} & =\dint \dfrac{d^{D}q}{i\pi ^{\frac{D}{2}}}%
\dfrac{1}{\left( q^{2}-m^{2}\right) ^{a}}.%
\end{array}%
\end{equation}%
Performing the corresponding algebra allows us to obtain the required MRE:

\begin{equation}
\begin{array}{ll}
\dint \dfrac{d^{D}q}{i\pi ^{\frac{D}{2}}}\dfrac{1}{\left( q^{2}-m^{2}\right)
^{a}} & =\dfrac{1}{\Gamma (a)}\dint\limits_{0}^{\infty }dx\;x^{a-1}\exp
(xm^{2})\dint \dfrac{d^{D}q}{i\pi ^{\frac{D}{2}}}\;\exp (-xq^{2}) \\
&  \\
& =\dfrac{(-1)^{-\frac{D}{2}}}{\Gamma (a)}\dsum\limits_{n}\phi _{n}\;\left(
-m^{2}\right) ^{n}\left\langle a-\frac{D}{2}+n\right\rangle .%
\end{array}%
\end{equation}%
Pictorially this is summarized in the following way:

\begin{equation}
\fbox{$%
\begin{array}{c}
a,m \\
\begin{minipage}{2.3cm} \includegraphics[scale=.7]
{graph_C.eps} \end{minipage} \\
\end{array}%
=\dfrac{(-1)^{-\frac{D}{2}}}{\Gamma (a)}\dsum\limits_{n}\phi _{n}\ \left(
-m^{2}\right) ^{n}\left\langle a-\tfrac{D}{2}+n\right\rangle $}  \label{ff32}
\end{equation}

\paragraph{Alternative II : Using the MRE of the massive propagator}

An alternative form for finding the MRE of this module can be implemented
using the MRE of the denominator in the loop integral, which extracts the
mass out of the integral and leaves the equivalent of a vacuum massless
bubble. Meanwhile, the corresponding MRE of a massive propagator can be
easily found using the fundamental equation $\left( \ref{f22}\right) $, that
is:

\begin{equation}
\frac{1}{\left( q^{2}-m^{2}\right) ^{a}}=\dsum\limits_{n_{1},n_{2}}\phi
_{n_{1},n_{2}}\;\left( -m^{2}\right) ^{n_{1}}\left( q^{2}\right) ^{n_{2}}%
\frac{\left\langle a+n_{1}+n_{2}\right\rangle }{\Gamma (a)}.
\end{equation}%
We can now rewrite the following equation for a massive bubble:

\begin{equation}
\begin{array}{ccc}
a,m &  & -n_{2} \\
\begin{minipage}{2.3cm} \includegraphics[scale=.7]
{graph_C.eps} \end{minipage} & =\dsum\limits_{n_{1},n_{2}}\phi
_{n_{1},n_{2}}\;\left( -m^{2}\right) ^{n_{1}}\dfrac{\left\langle
a+n_{1}+n_{2}\right\rangle }{\Gamma (a)}\;\times &
\begin{minipage}{2.3cm} \includegraphics[scale=.7]
{graph_C.eps} \end{minipage}%
\end{array}%
\end{equation}%
where the massless bubble can be evaluated using equation $\left( \ref{ff31}%
\right) $, and then we finally obtain the MRE for this case:

\begin{equation}
\fbox{$%
\begin{array}{c}
a,m \\
\begin{minipage}{2.3cm} \includegraphics[scale=.7]
{graph_C.eps} \end{minipage} \\
\end{array}%
=\dfrac{(-1)^{-\frac{D}{2}}}{\Gamma (a)}\dsum\limits_{n_{1},n_{2}}\phi
_{n_{1},n_{2}}\ \left( -m^{2}\right) ^{n_{1}}\dfrac{\left\langle
a+n_{1}+n_{2}\right\rangle \left\langle -n_{2}-\tfrac{D}{2}\right\rangle }{%
\Gamma (-n_{2})}$}  \label{ff33}
\end{equation}%
Although $\left( \ref{ff32}\right) $ and $\left( \ref{ff33}\right) $ are
equivalent, it is clear that $\left( \ref{ff32}\right) $ is a more compact
result, which makes actual evaluation easier.

\subsection{MRE of a bubble with two propagators}

\subsubsection{Trivial cases : $\overline{G}_{A}$, $\overline{G}_{C}$}

There are two cases of vacuum bubbles with two propagators which are
reducible to one propagator. This happens when the two propagators which
form the bubble have equal mass parameters, in which case it is possible to
use the pictorial equation $\left( \ref{ff4}\right) $ in order to simplify
the graph.

The first case corresponds to a composition of two massless propagators,
which results in an expression for the $1$-$loop$ function $\overline{G}_{A}$
and which is reduced as follows:

\begin{equation}
\overline{G}_{A}(a_{1},a_{2})=%
\begin{array}{ccc}
a_{1} &  & (a_{1}+a_{2}) \\
\begin{minipage}{2.5cm} \includegraphics[scale=.7]
{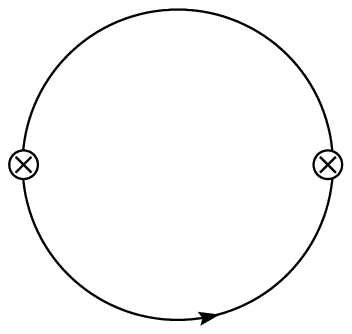} \end{minipage} & = &
\begin{minipage}{2.3cm} \includegraphics[scale=.7]
{graph_C.eps} \end{minipage}. \\
a_{2} &  &
\end{array}%
\end{equation}%
With the help of equation $\left( \ref{ff31}\right) $ one then obtains the
following expression for $\overline{G}_{A}$:

\begin{equation}
\overline{G}_{A}(a_{1},a_{2})=%
\begin{array}{c}
a_{1} \\
\begin{minipage}{2.5cm} \includegraphics[scale=.7]
{graph_B.eps} \end{minipage} \\
a_{2}%
\end{array}%
=(-1)^{-\frac{D}{2}}\dfrac{\left\langle a_{1}+a_{2}-\tfrac{D}{2}%
\right\rangle }{\Gamma (a_{1}+a_{2})}.
\end{equation}%
Analogously, for the case of two propagators with the same arbitrary mass $m$%
, the resulting $1$-$loop$ function describes $\overline{G}_{C}$:

\begin{equation}
\overline{G}_{C}\left( a_{1},a_{2};m^{2}\right) =%
\begin{array}{ccc}
a_{1},m &  & (a_{1}+a_{2}),m \\
\begin{minipage}{2.5cm} \includegraphics[scale=.7]
{graph_B.eps} \end{minipage} & = &
\begin{minipage}{2.3cm} \includegraphics[scale=.7]
{graph_C.eps} \end{minipage}, \\
a_{2},m &  &
\end{array}%
\end{equation}%
and then using $\left( \ref{ff32}\right) $ we get the MRE for this module:

\begin{equation}
\overline{G}_{C}\left( a_{1},a_{2};m^{2}\right) =%
\begin{array}{c}
a_{1},m \\
\begin{minipage}{2.5cm} \includegraphics[scale=.7]
{graph_B.eps} \end{minipage} \\
a_{2},m%
\end{array}%
=\dfrac{(-1)^{-\frac{D}{2}}}{\Gamma (a_{1}+a_{2})}\dsum\limits_{n}\phi _{n}\
\left( -m^{2}\right) ^{n}\left\langle a_{1}+a_{2}-\frac{D}{2}+n\right\rangle
.
\end{equation}

\subsubsection{One massive propagator : $\overline{G}_{B}$}

This case defines the loop function $\overline{G}_{B}$ and just as was done
before we will consider two different possible forms to represent the MRE.
This module is given by the integral representation:

\begin{equation}
\overline{G}_{B}(a_{1},a_{2};m^{2})=%
\begin{array}{c}
a_{1} \\
\begin{minipage}{2.5cm} \includegraphics[scale=.7]
{graph_B.eps} \end{minipage} \\
a_{2},m%
\end{array}%
=\dint \dfrac{d^{D}q}{i\pi ^{\frac{D}{2}}}\dfrac{1}{\left( q^{2}\right)
^{a_{1}}\left( q^{2}-m^{2}\right) ^{a_{2}}}.
\end{equation}

\paragraph{Alternative I : Direct parametrization of the integral}

The parametrization and subsequent evaluation of the loop integral gives us:

\begin{equation}
\begin{array}{ll}
\dint \dfrac{d^{D}q}{i\pi ^{\frac{D}{2}}}\dfrac{1}{\left( q^{2}\right)
^{a_{1}}\left( q^{2}-m^{2}\right) ^{a_{2}}} & =\dfrac{1}{\Gamma
(a_{1})\Gamma (a_{2})}\dint\limits_{0}^{\infty
}dxdy\;x^{a_{1}-1}y^{a_{2}-1}\exp (y\,m^{2})\dint \dfrac{d^{D}q}{i\pi ^{%
\frac{D}{2}}}\;\exp \left[ -(x+y)q^{2}\right] \\
&  \\
& =\dfrac{(-1)^{-\frac{D}{2}}}{\Gamma (a_{1})\Gamma (a_{2})}%
\dint\limits_{0}^{\infty }dxdy\;x^{a_{1}-1}y^{a_{2}-1}\dfrac{\exp (y\,m^{2})%
}{(x+y)^{\frac{D}{2}}}.%
\end{array}%
\end{equation}%
After some algebra we get the MRE of this case:

\begin{equation}
\fbox{$\overline{G}_{B}(a_{1},a_{2};m^{2})=%
\begin{array}{c}
a_{1} \\
\begin{minipage}{2.5cm} \includegraphics[scale=.7]
{graph_B.eps} \end{minipage} \\
a_{2},m%
\end{array}%
=\dfrac{(-1)^{-\frac{D}{2}}}{\Gamma (a_{1})}\dsum\limits_{n_{1},n_{2},n_{3}}%
\phi _{n_{1},n_{2},n_{3}}\ \left( -m^{2}\right) ^{n_{1}}\dfrac{\Delta
_{1}\Delta _{2}\Delta _{3}}{\Gamma (\frac{D}{2})}$}
\end{equation}%
where:

\begin{equation}
\left\{
\begin{array}{l}
\Delta _{1}=\left\langle \frac{D}{2}+n_{2}+n_{3}\right\rangle , \\
\Delta _{2}=\left\langle a_{1}+n_{2}\right\rangle , \\
\Delta _{3}=\left\langle a_{2}+n_{1}+n_{3}\right\rangle .%
\end{array}%
\right.
\end{equation}

\paragraph{Alternative II : Using the MRE of the massive propagator :}

The second alternative implies to extract the mass of the loop integral,
that is, to find the MRE of the massive propagator. Then we have:

\begin{equation}
\begin{array}{cccc}
& a_{1} &  & \left( a_{1}-n_{2}\right) \\
\overline{G}_{B}(a_{1},a_{2};m^{2})= &
\begin{minipage}{2.5cm} \includegraphics[scale=.7]
{graph_B.eps} \end{minipage} & =\dsum\limits_{n_{1},n_{2}}\phi
_{n_{1},n_{2}}\;\left( -m^{2}\right) ^{n_{1}}\dfrac{\left\langle
a_{2}+n_{1}+n_{2}\right\rangle }{\Gamma (a_{2})}\;\times &
\begin{minipage}{2.3cm} \includegraphics[scale=.7]
{graph_C.eps} \end{minipage} \\
& a_{2},m &  &
\end{array}%
\end{equation}%
or equivalently:

\begin{equation}
\fbox{$%
\begin{array}{ccc}
& a_{1} &  \\
\overline{G}_{B}(a_{1},a_{2};m^{2})= &
\begin{minipage}{2.5cm} \includegraphics[scale=.7]
{graph_B.eps} \end{minipage} & =\dfrac{(-1)^{-\frac{D}{2}}}{\Gamma (a_{2})}%
\dsum\limits_{n_{1},n_{2}}\phi _{n_{1},n_{2}}\ \left( -m^{2}\right) ^{n_{1}}%
\dfrac{\left\langle a_{2}+n_{1}+n_{2}\right\rangle \left\langle a_{1}-n_{2}-%
\frac{D}{2}\right\rangle }{\Gamma (a_{1}-n_{2})} \\
& a_{2},m &
\end{array}%
$}  \label{ff35}
\end{equation}%
Is evident that equation $\left( \ref{ff35}\right) $\ is the most compact
for function $\overline{G}_{B}$.

\subsubsection{Propagators with different masses : $\overline{G}_{D}$}

The next function is associated to a bubble with two propagators of
different mass, which defines the loop function $\overline{G}_{D}.$ Again we
will deduce this function in two alternative ways, as was done before:

\paragraph{Alternative I : Expansion of the complete integral}

Applying Schwinger's parametrization allows to find the following structure
for this module:

\begin{equation}
\begin{array}{cc}
a_{1},M &  \\
\begin{minipage}{2.5cm} \includegraphics[scale=.7]
{graph_B.eps} \end{minipage} & =\dfrac{1}{\Gamma (a_{1})\Gamma (a_{2})}%
\dint\limits_{0}^{\infty }dxdy\;x^{a_{1}-1}y^{a_{2}-1}\exp (x\,M^{2})\exp
(y\,m^{2})\dint \dfrac{d^{D}q}{i\pi ^{\frac{D}{2}}}\;\exp \left[ -(x+y)q^{2}%
\right] . \\
a_{2},m &
\end{array}%
\end{equation}%
Performing the evaluations and replacements already described, we get the
MRE for this case:

\begin{equation}
\fbox{$\overline{G}_{D}(a_{1},a_{2};M^{2},m^{2})=%
\begin{array}{c}
a_{1},M \\
\begin{minipage}{2.5cm} \includegraphics[scale=.7]
{graph_B.eps} \end{minipage} \\
a_{2},m%
\end{array}%
=\dfrac{(-1)^{-\frac{D}{2}}}{\Gamma (a_{1})\Gamma (a_{2})}\dsum\limits
_{\substack{ n_{1},n_{2}  \\ n_{3},n_{4}}}\phi _{n_{1},..,n_{4}}\ \left(
-m^{2}\right) ^{n_{1}}\left( -M^{2}\right) ^{n_{2}}\dfrac{\Delta _{1}\Delta
_{2}\Delta _{3}}{\Gamma (\frac{D}{2})}$}
\end{equation}%
where the associated constraints are:

\begin{equation}
\left\{
\begin{array}{l}
\Delta _{1}=\left\langle \frac{D}{2}+n_{3}+n_{4}\right\rangle , \\
\Delta _{2}=\left\langle a_{1}+n_{1}+n_{3}\right\rangle , \\
\Delta _{3}=\left\langle a_{2}+n_{2}+n_{4}\right\rangle .%
\end{array}%
\right.
\end{equation}

\paragraph{Alternative II : Expansion of massive propagators}

The momentum integral is given by:

\begin{equation}
\overline{G}_{D}(a_{1},a_{2};M^{2},m^{2})=%
\begin{array}{c}
a_{1},M \\
\begin{minipage}{2.5cm} \includegraphics[scale=.7]
{graph_B.eps} \end{minipage} \\
a_{2},m%
\end{array}%
=\dint \dfrac{d^{D}q}{i\pi ^{\frac{D}{2}}}\dfrac{1}{\left(
q^{2}-M^{2}\right) ^{a_{1}}\left( q^{2}-m^{2}\right) ^{a_{2}}},
\end{equation}%
then we expand each massive propagator:

\begin{equation}
\begin{array}{ll}
\begin{array}{c}
a_{1},M \\
\begin{minipage}{2.5cm} \includegraphics[scale=.7]
{graph_B.eps} \end{minipage} \\
a_{2},m%
\end{array}%
= & \dsum\limits_{n_{1},n_{2}}\phi _{n_{1},n_{2}}\;\left( -M^{2}\right)
^{n_{1}}\dfrac{\left\langle a_{1}+n_{1}+n_{2}\right\rangle }{\Gamma (a_{1})}%
\dsum\limits_{n_{3},n_{4}}\phi _{n_{3},n_{4}}\;\left( -m^{2}\right) ^{n_{3}}%
\dfrac{\left\langle a_{2}+n_{3}+n_{4}\right\rangle }{\Gamma (\nu _{2})} \\
&  \\
& \times
\begin{array}{c}
\left( -n_{2}-n_{4}\right) \\
\begin{minipage}{2.3cm} \includegraphics[scale=.7]{graph_C.eps}
\end{minipage} \\
\end{array}%
\end{array}%
\end{equation}%
andDIDO using formula $\left( \ref{ff31}\right) $ we obtain the loop
function $\overline{G}_{D}$:

\begin{equation}
\fbox{$%
\begin{array}{ll}
\overline{G}_{D}(a_{1},a_{2};M^{2},m^{2}) & =%
\begin{array}{c}
a_{1},M \\
\begin{minipage}{2.5cm} \includegraphics[scale=.7]
{graph_B.eps} \end{minipage} \\
a_{2},m%
\end{array}
\\
&  \\
& =\dfrac{(-1)^{-\frac{D}{2}}}{\Gamma (a_{1})\Gamma (a_{2})}\dsum\limits
_{\substack{ n_{1},n_{2}  \\ n_{3},n_{4}}}\phi _{n_{1},..,n_{4}}\;\left(
-M^{2}\right) ^{n_{1}}\left( -m^{2}\right) ^{n_{3}}\dfrac{\Delta _{1}\Delta
_{2}\Delta _{3}}{\Gamma (-n_{2}-n_{4})}%
\end{array}%
$}
\end{equation}%
with the following constraints:

\begin{equation}
\left\{
\begin{array}{l}
\Delta _{1}=\left\langle a_{1}+n_{1}+n_{2}\right\rangle , \\
\Delta _{2}=\left\langle a_{2}+n_{3}+n_{4}\right\rangle , \\
\Delta _{3}=\left\langle -n_{2}-n_{4}-\frac{D}{2}\right\rangle .%
\end{array}%
\right.
\end{equation}%
In all the massive cases we can use two equivalent ways of finding the MRE
of the present module, but one of them will produce a mathematical expansion
in a more reduced or minimal form. More compact MREs allow a considerable
reduction of the possibilities of evaluating the sums of the expansion with
the available Kronecker deltas. Remember that in the case of the MRE of a
certain Feynman diagram, each alternative way of doing these evaluations
generates a term (hypergeometric series) of the final solution, and
therefore a minimal MRE allows to eliminate irrelevant or non existing terms
(cases where it is not possible mathematically to perform the sum using the
Kronecker deltas).

\section{Applications}

\qquad In this section we will use the previously obtained $1$-$loop$
functions and present three examples that will show and explain the IBFE
methodology applied modularly. Together with this we will be able to
visualize the advantages of this procedure compared with IBFE applied to the
Schwinger's parametric representation of the complete diagram. In our
previous work \cite{IGoIBFE}, we show as find explicitly solutions to start
of the MRE equation of a Feynman diagram. In this work, we just compare the
ways to obtain the respective MRE associated to a topology.

\subsection{Example I : Radiative correction to the three loop massless
propagator}

Let us consider as a first example the three loop correction to the massless
propagator, with momentum $p$, as shown in the graphical formula $\left( \ref%
{ff26}\right) $, applying the fractional expansion method loop by loop, and
using the $1$-$loop$ functions previously obtained. In this case it is clear
that we need only $1$-$loop$ functions of type $G_{A}$.

\begin{equation}
G=%
\begin{minipage}{4.7cm} \includegraphics[scale=.7]
{ex0_1.eps} \end{minipage}  \label{ff26}
\end{equation}%
We start by reducing the two bubble insertions directly, using the pictorial
formula $\left( \ref{ff3}\right) $, which allows us to get:

\begin{equation}
\begin{array}{cc}
& \left( a_{5}-n_{1}\right) \\
G=\dsum\limits_{n_{1},n_{2}}G_{A}\left( a_{1},a_{2};n_{1}\right) \times
G_{A}\left( a_{3},a_{4};n_{2}\right) \;\times &
\begin{minipage}{4.5cm} \includegraphics[scale=.7]
{graph_A.eps} \end{minipage} \\
& \left( a_{6}-n_{2}\right)%
\end{array}%
\end{equation}%
We can use $\left( \ref{ff4}\right) $ to sum the propagator indices, and
then apply formula $\left( \ref{ff3}\right) $ for finally getting the
diagram MRE in terms of a sum of products of $1$-$loop$ functions $G_{A}$:

\begin{equation}
\begin{array}{cc}
& -n_{3} \\
G=\dsum\limits_{n_{1},..,n_{3}}G_{A}\left( a_{1},a_{2};n_{1}\right) \times
G_{A}\left( a_{3},a_{4};n_{2}\right) \times G_{A}\left(
a_{5}-n_{1},a_{6}-n_{2};n_{3}\right) \;\times &
\begin{minipage}{3.3cm} \includegraphics[scale=.8]
{graph_D.eps} \end{minipage} \\
&
\end{array}%
\end{equation}%
This result can be written also as:

\begin{equation}
G=\dsum\limits_{n_{1},..,n_{3}}G_{A}\left( a_{1},a_{2};n_{1}\right) \times
G_{A}\left( a_{3},a_{4};n_{2}\right) \times G_{A}\left(
a_{5}-n_{1},a_{6}-n_{2};n_{3}\right) \;(p^{2})^{n_{3}},  \label{ff6}
\end{equation}%
where the $1$-$loop$ functions $G_{A}$ for each case are determined,
according to equation $\left( \ref{ff5}\right) $, by the following
expression:

\begin{equation}
G_{A}\left( a_{1},a_{2};n_{1}\right) =\dfrac{(-1)^{-\frac{D}{2}}}{\Gamma
(a_{1})\Gamma (a_{2})}\dsum\limits_{n_{4},n_{5}}\phi _{n_{1},n_{4},n_{5}}\;%
\dfrac{\Delta _{1}\Delta _{2}\Delta _{3}}{\Gamma (\frac{D}{2}+n_{1})},
\end{equation}%
with constraints that are given by the identities:

\begin{equation}
\left\{
\begin{array}{l}
\Delta _{1}=\left\langle \frac{D}{2}+n_{1}+n_{4}+n_{5}\right\rangle , \\
\Delta _{2}=\left\langle a_{1}+n_{1}+n_{4}\right\rangle , \\
\Delta _{3}=\left\langle a_{2}+n_{1}+n_{5}\right\rangle .%
\end{array}%
\right.
\end{equation}%
Similarly one has:

\begin{equation}
G_{A}\left( a_{3},a_{4};n_{2}\right) =\dfrac{(-1)^{-\frac{D}{2}}}{\Gamma
(a_{3})\Gamma (a_{4})}\dsum\limits_{n_{6},n_{7}}\phi _{n_{2},n_{6},n_{7}}\;%
\dfrac{\Delta _{4}\Delta _{5}\Delta _{6}}{\Gamma (\frac{D}{2}+n_{2})},
\end{equation}

\begin{equation}
\left\{
\begin{array}{l}
\Delta _{4}=\left\langle \frac{D}{2}+n_{2}+n_{6}+n_{7}\right\rangle , \\
\Delta _{5}=\left\langle a_{3}+n_{2}+n_{6}\right\rangle , \\
\Delta _{6}=\left\langle a_{4}+n_{2}+n_{7}\right\rangle ,%
\end{array}%
\right.
\end{equation}%
and finally:

\begin{equation}
G_{A}\left( a_{5}-n_{1},a_{6}-n_{2};n_{3}\right) =\dfrac{(-1)^{-\frac{D}{2}}%
}{\Gamma (a_{5}-n_{1})\Gamma (a_{6}-n_{2})}\dsum\limits_{n_{8},n_{9}}\phi
_{n_{3},n_{8},n_{9}}\;\dfrac{\Delta _{7}\Delta _{8}\Delta _{9}}{\Gamma (%
\frac{D}{2}+n_{3})},
\end{equation}

\begin{equation}
\left\{
\begin{array}{l}
\Delta _{7}=\left\langle \frac{D}{2}+n_{3}+n_{8}+n_{9}\right\rangle , \\
\Delta _{8}=\left\langle a_{5}-n_{1}+n_{3}+n_{8}\right\rangle , \\
\Delta _{9}=\left\langle a_{6}-n_{2}+n_{3}+n_{9}\right\rangle .%
\end{array}%
\right.
\end{equation}%
Replacing these series in equation $\left( \ref{ff6}\right) $ allows to get
the diagram MRE:

\begin{equation}
\fbox{$G=\dfrac{(-1)^{-\frac{3D}{2}}}{\prod\nolimits_{i=1}^{4}\Gamma (a_{i})}%
\dsum\limits_{n_{1},..,n_{9}}\phi _{n_{1},..,n_{9}}$\ $\dfrac{(p^{2})^{n_{3}}%
}{\Gamma (\frac{D}{2}+n_{1})\Gamma (\frac{D}{2}+n_{2})\Gamma (\frac{D}{2}%
+n_{3})}\dfrac{\prod\nolimits_{j=1}^{9}\Delta _{j}}{\Gamma
(a_{5}-n_{1})\Gamma (a_{6}-n_{2})}$}  \label{ff19}
\end{equation}%
In order to show the advantages of modular IBFE, we show in Table I a
comparison of the MRE obtained in this form with respect to that one that
evaluates the complete diagram, that is which uses the parametric
representation that incudes all loops simultaneously.

\begin{equation}
\begin{tabular}{lll}
\hline
& Complete (better factorization) & Modular \\ \hline
Multiplicity multiregion series $\left( {\sigma }\right) $ &
\multicolumn{1}{c}{11} & \multicolumn{1}{c}{9} \\
Kronecker deltas of the expansion $\left( {\delta }\right) $ &
\multicolumn{1}{c}{11} & \multicolumn{1}{c}{9} \\
Multiplicity resulting series $\left( {\sigma -\delta }\right) $ &
\multicolumn{1}{c}{0} & \multicolumn{1}{c}{0} \\
Posible contributions to the solution $\left( {C}_{\delta }^{\sigma }\right)
$ & \multicolumn{1}{c}{1} & \multicolumn{1}{c}{1} \\ \hline
\end{tabular}
\tag{$Table\ I$}
\end{equation}%
The expression $\left( \ref{ff19}\right) $ is equivalent to equation $\left( %
\ref{ff7}\right) $, which was deduced for the same diagram but using the
usual $1$-$loop$ function $\left( \ref{ff17}\right) $. The first conclusion
is that the fractional expansion technique is, for cases of loop by loop
reducible massless diagrams, more tedious than the conventional. In spite of
this, we will see that for cases in which the diagram contains different
mass scales, the present technique es very powerful, and this will become
clear in the following example.

\subsection{Example II : Propagator with two mass scales and two loops}

Let us consider a graph which consists of a two loop propagator with two
different mass scales, as shown in the figure:

\begin{equation}
\begin{minipage}{5.5cm} \includegraphics[scale=.7]{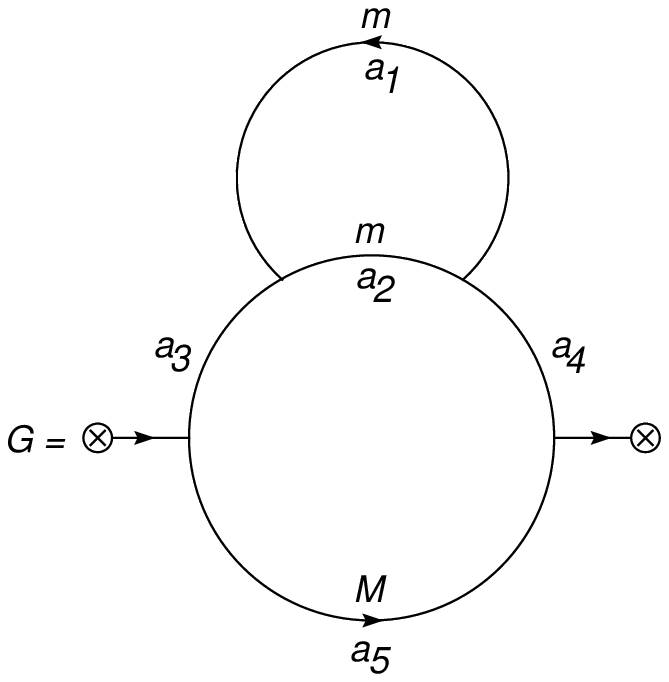} \end{minipage}
\end{equation}%
We first reduce the loop associated to the mass $m$, which uses the $1$-$%
loop $ function $G_{C}$ (equation $\left( \ref{ff8}\right) $), and then we
get:

\begin{equation}
\begin{array}{cc}
& \left( a_{3}+a_{4}-n_{1}\right) \\
G=\dsum\limits_{n_{1}}G_{C}\left( a_{1},a_{2};n_{1};m^{2}\right) \;\times &
\begin{minipage}{4.5cm} \includegraphics[scale=.7]
{graph_A.eps} \end{minipage} \\
& a_{5},M%
\end{array}%
\end{equation}%
For the resulting loop reduction we use formula $\left( \ref{ff9}\right) $,
obtaining the massive diagram MRE:

\begin{equation}
\begin{array}{cc}
& -n_{2} \\
G=\dsum\limits_{n_{1},n_{2}}G_{C}\left( a_{1},a_{2};n_{1};m^{2}\right)
\times G_{B}\left( a_{5},a_{3}+a_{4}-n_{1};n_{2};M^{2}\right) \;\times &
\begin{minipage}{3.3cm} \includegraphics[scale=.8]
{graph_D.eps} \end{minipage} \\
&
\end{array}%
\end{equation}%
which is equivalent to the algebraic expression:

\begin{equation}
G=\dsum\limits_{n_{1},n_{2}}G_{C}\left( a_{1},a_{2};n_{1};m^{2}\right)
\times G_{B}\left( a_{5},a_{3}+a_{4}-n_{1};n_{2};M^{2}\right) \;\left(
p^{2}\right) ^{n_{2}}.  \label{ff10}
\end{equation}%
The $1$-$loop$ functions $G_{C}$ and $G_{B}$ in $\left( \ref{ff10}\right) $
are given by:

\begin{equation}
G_{C}\left( a_{1},a_{2};n_{1};m^{2}\right) =\dfrac{(-1)^{-\frac{D}{2}}}{%
\Gamma (a_{1})\Gamma (a_{2})}\dsum\limits_{n_{3},n_{4},n_{5}}\phi
_{n_{1},n_{3},n_{4},n_{5}}\;(-m^{2})^{n_{3}}\frac{\Delta _{1}\Delta
_{2}\Delta _{3}}{\Gamma (\frac{D}{2}+n_{1}-n_{3})},
\end{equation}%
where the constraints are:

\begin{equation}
\left\{
\begin{array}{l}
\Delta _{1}=\left\langle \frac{D}{2}+n_{1}-n_{3}+n_{4}+n_{5}\right\rangle ,
\\
\Delta _{2}=\left\langle a_{1}+n_{1}+n_{4}\right\rangle , \\
\Delta _{3}=\left\langle a_{2}+n_{1}+n_{5}\right\rangle ,%
\end{array}%
\right.
\end{equation}%
and:

\begin{equation}
G_{B}(a_{5},a_{3}+a_{4}-n_{1};n_{2};M^{2})=\dfrac{(-1)^{-\frac{D}{2}}}{%
\Gamma (a_{5})\Gamma (a_{3}+a_{4}-n_{1})}\dsum\limits_{n_{6},n_{7},n_{8}}%
\phi _{n_{2},n_{6},n_{7},n_{8}}\;(-M^{2})^{n_{6}}\dfrac{\Delta _{4}\Delta
_{5}\Delta _{6}}{\Gamma (\frac{D}{2}+n_{2})},
\end{equation}%
with the constraints:

\begin{equation}
\left\{
\begin{array}{l}
\Delta _{4}=\left\langle \frac{D}{2}+n_{2}+n_{7}+n_{8}\right\rangle , \\
\Delta _{5}=\left\langle a_{5}+n_{2}+n_{6}+n_{7}\right\rangle , \\
\Delta _{6}=\left\langle a_{3}+a_{4}-n_{1}+n_{2}+n_{8}\right\rangle .%
\end{array}%
\right.
\end{equation}%
The replacement in $\left( \ref{ff10}\right) $ allows to finally get the MRE
of diagram $G$:

\begin{equation}
\fbox{$G=\dfrac{(-1)^{-D}}{\Gamma (a_{1})\Gamma (a_{2})\Gamma (a_{5})}%
\dsum\limits_{n_{1},..,n_{8}}\phi _{n_{1},..,n_{8}}$\ $\dfrac{\left(
p^{2}\right) ^{n_{2}}(-m^{2})^{n_{3}}(-M^{2})^{n_{6}}}{\Gamma (\frac{D}{2}%
+n_{1}-n_{3})\Gamma (\frac{D}{2}+n_{2})}\dfrac{\Delta _{1}\Delta _{2}\Delta
_{3}\Delta _{4}\Delta _{5}\Delta _{6}}{\Gamma (a_{3}+a_{4}-n_{1})}$}
\end{equation}%
Operationally it is not possible to apply directly the conventional $1$-$%
loop $ reduction $\left( \ref{ff17}\right) $ to this massive example, as has
been done here with the fractionally expanded $1$-$loop$ functions.
Moreover, it is also possible to verify that the diagram MRE obtained loop
by loop is far more direct and simple than to get this expansion considering
all loops simultaneously. This can be seen from Table II, where we observe
that the modular application of the IBFE technique not only reduces the MRE
of the diagram in terms of sums and deltas, but also brings in a significant
reduction of the number of irrelevant terms and therefore the calculation
time of the solution is optimized.

\begin{equation}
\begin{tabular}{lll}
\hline
& Complete (better factorization) & Modular \\ \hline
Multiplicity multiregion series $\left( {\sigma }\right) $ &
\multicolumn{1}{c}{11} & \multicolumn{1}{c}{8} \\
Kronecker deltas of the expansion $\left( {\delta }\right) $ &
\multicolumn{1}{c}{9} & \multicolumn{1}{c}{6} \\
Multiplicity resulting series $\left( {\sigma -\delta }\right) $ &
\multicolumn{1}{c}{2} & \multicolumn{1}{c}{2} \\
Possible contributions to the solution $\left( {C}_{\delta }^{\sigma
}\right) $ & \multicolumn{1}{c}{55} & \multicolumn{1}{c}{28} \\ \hline
\end{tabular}
\tag{$Tabla\ II$}
\end{equation}

\subsection{Example III : Vacuum fluctuation with three mass scales and five
loops}

The next diagram that we will analyze is one without external lines,
composed of five loops and three different mass scales $\left\{
m_{1},m_{2},M\right\} $, arranged as shown here:

\begin{equation}
G=\left[ \begin{minipage}{7cm} \includegraphics[scale=.7]{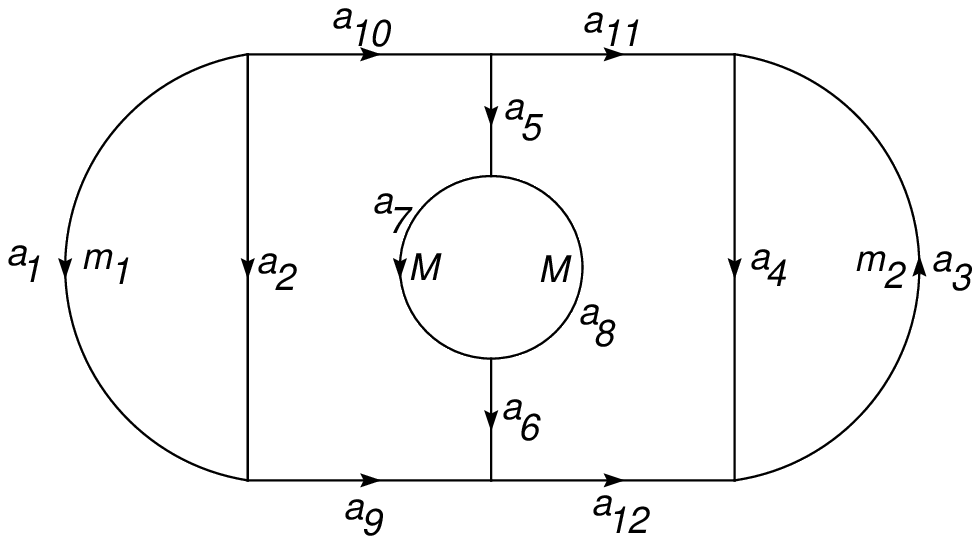}
\end{minipage}\right]
\end{equation}%
In the first step we reduce the left and right massive loops in terms of the
$1$-$loop$ functions $G_{B}$:%
\begin{equation}
G=\dsum\limits_{n_{1},n_{2}}G_{B}(a_{1},a_{2};n_{1};m_{1}^{2})\times
G_{B}(a_{3},a_{4};n_{2};m_{2}^{2})\;\times \left[ \begin{minipage}{7.2cm}
\includegraphics[scale=.7]{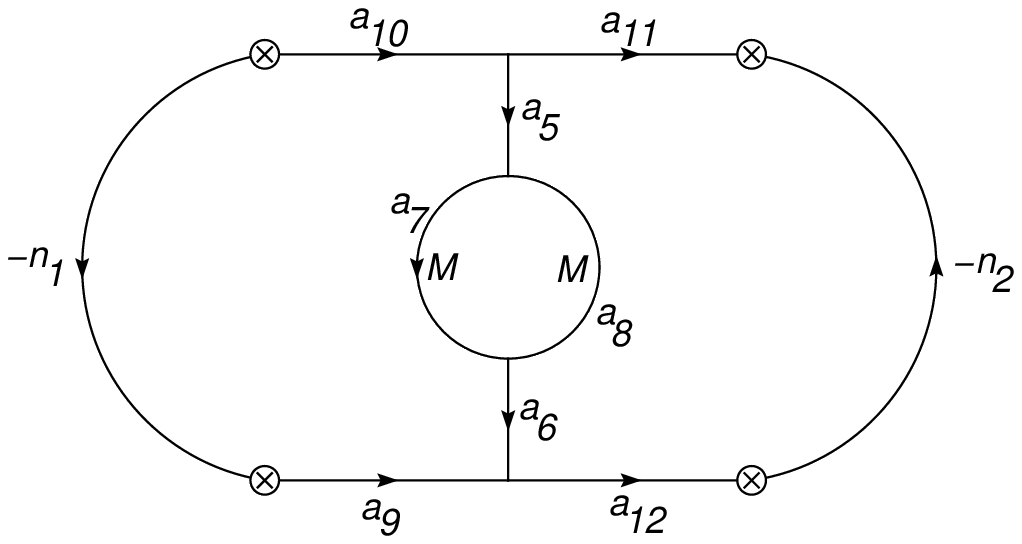} \end{minipage}\right]
\end{equation}%
Applying formula $\left( \ref{ff4}\right) $ in order to sum the propagator
indices, we get the equivalent equation:

\begin{equation}
\begin{array}{ll}
G= & \dsum\limits_{n_{1},n_{2}}G_{B}(a_{1},a_{2};n_{1};m_{1}^{2})\times
G_{B}(a_{3},a_{4};n_{2};m_{2}^{2}) \\
& \times \left[ \begin{minipage}{9.7cm}
\includegraphics[scale=.7]{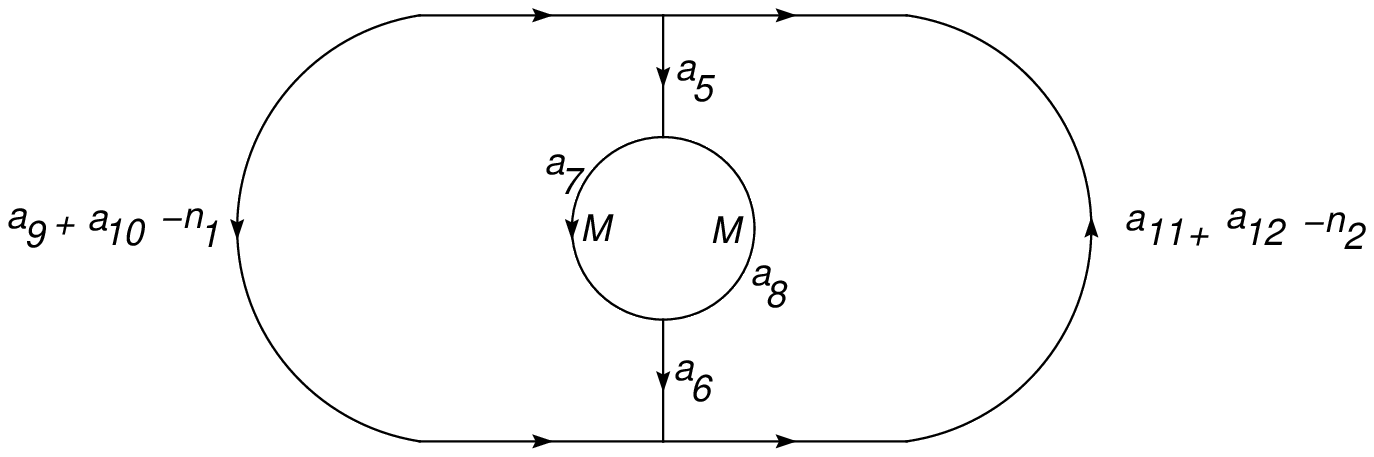} \end{minipage}\right]%
\end{array}
\notag
\end{equation}%
Then we eliminate the internal bubble:

\begin{equation}
\begin{array}{ll}
G= & \dsum\limits_{n_{1},..,n_{3}}G_{B}(a_{1},a_{2};n_{1};m_{1}^{2})\times
G_{B}(a_{3},a_{4};n_{2};m_{2}^{2})\times G_{C}(a_{7},a_{8};n_{3};M^{2}) \\
& \times \left[ \begin{minipage}{9.7cm}
\includegraphics[scale=.7]{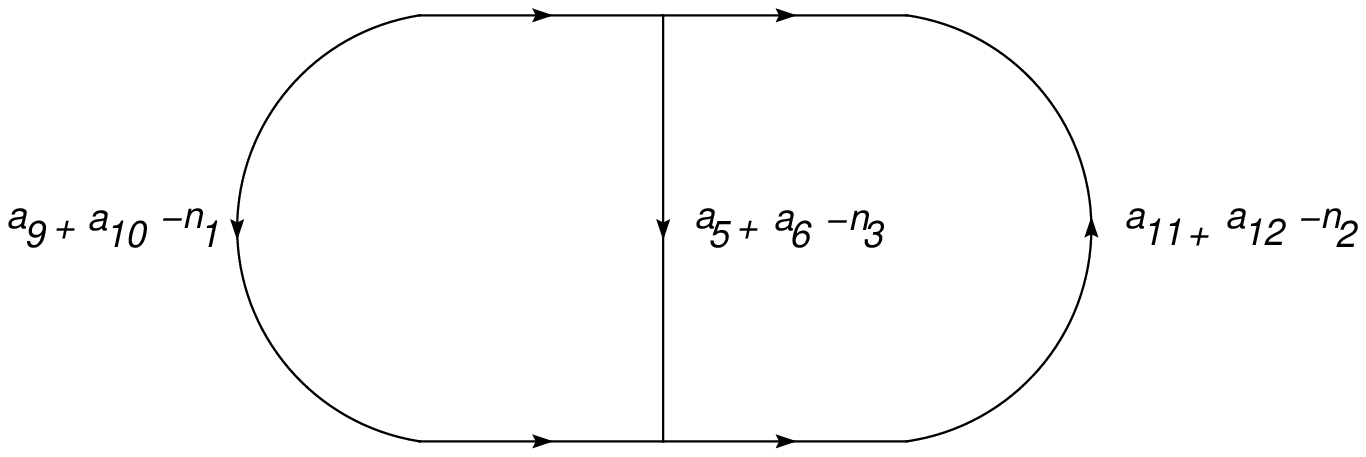} \end{minipage}\right]%
\end{array}
\notag
\end{equation}%
continuing with a reduction of the left loop:

\begin{equation}
\begin{array}{ll}
G= & \dsum\limits_{n_{1},..,n_{4}}G_{B}(a_{1},a_{2};n_{1};m_{1}^{2})\times
G_{B}(a_{3},a_{4};n_{2};m_{2}^{2})\times G_{C}(a_{7},a_{8};n_{3};M^{2}) \\
&  \\
& G_{A}(a_{9,10}-n_{1},a_{5,6}-n_{3};n_{4})\times \left[ -n_{4}%
\begin{minipage}{2.5cm} \includegraphics[scale=.7]{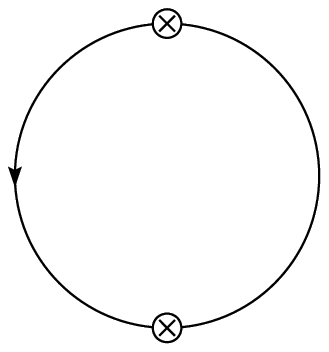}
\end{minipage}\;a_{11}+a_{12}-n_{2}\right]%
\end{array}%
\end{equation}%
The last loop reduction is represented by the function $\overline{G}_{A}$,
which determines finally the MRE of the diagram:

\begin{equation}
\begin{array}{ll}
G= & \dsum\limits_{n_{1},..,n_{4}}G_{B}(a_{1},a_{2};n_{1};m_{1}^{2})\times
G_{B}(a_{3},a_{4};n_{2};m_{2}^{2})\times
G_{C}(a_{7},a_{8};n_{3};M^{2})\times
G_{A}(a_{9,10}-n_{1},a_{5,6}-n_{3};n_{4}) \\
&  \\
& \times \;\overline{G}_{A}(a_{11,12}-n_{2},-n_{4}).%
\end{array}%
\end{equation}%
The corresponding $1$-$loop$ functions in this example are given by:

\begin{equation}
G_{B}(a_{1},a_{2};n_{1};m_{1}^{2})=\dfrac{(-1)^{-\frac{D}{2}}}{\Gamma
(a_{1})\Gamma (a_{2})}\dsum\limits_{n_{5},..,n_{7}}\phi
_{n_{1},n_{5},..,n_{7}}\;\frac{(-m_{1}^{2})^{n_{7}}}{\Gamma (\frac{D}{2}%
+n_{1})}\Delta _{1}\Delta _{2}\Delta _{3},
\end{equation}

\begin{equation}
\left\{
\begin{array}{l}
\Delta _{1}=\left\langle a_{1}+n_{1}+n_{5}+n_{7}\right\rangle , \\
\Delta _{2}=\left\langle a_{2}+n_{1}+n_{6}\right\rangle , \\
\Delta _{3}=\left\langle \frac{D}{2}+n_{1}+n_{5}+n_{6}\right\rangle ,%
\end{array}%
\right.
\end{equation}

\begin{equation}
G_{B}(a_{3},a_{4};n_{2};m_{2}^{2})=\dfrac{(-1)^{-\frac{D}{2}}}{\Gamma
(a_{3})\Gamma (a_{4})}\dsum\limits_{n_{8},..,n_{10}}\phi
_{n_{2},n_{8},..,n_{10}}\;\dfrac{(-m_{2}^{2})^{n_{10}}}{\Gamma (\frac{D}{2}%
+n_{2})}\Delta _{4}\Delta _{5}\Delta _{6},
\end{equation}

\begin{equation}
\left\{
\begin{array}{l}
\Delta _{4}=\left\langle a_{3}+n_{2}+n_{8}+n_{10}\right\rangle , \\
\Delta _{5}=\left\langle a_{4}+n_{2}+n_{9}\right\rangle , \\
\Delta _{6}=\left\langle \frac{D}{2}+n_{2}+n_{8}+n_{9}\right\rangle ,%
\end{array}%
\right.
\end{equation}

\begin{equation}
G_{C}(a_{7},a_{8};n_{3};M^{2})=\dfrac{(-1)^{-\frac{D}{2}}}{\Gamma
(a_{7})\Gamma (a_{8})}\dsum\limits_{n_{11},..,n_{13}}\phi
_{n_{3},n_{11},..,n_{13}}\;\dfrac{(-M^{2})^{n_{13}}}{\Gamma (\frac{D}{2}%
+n_{3}-n_{13})}\Delta _{7}\Delta _{8}\Delta _{9},
\end{equation}

\begin{equation}
\left\{
\begin{array}{l}
\Delta _{7}=\left\langle a_{7}+n_{3}+n_{11}\right\rangle , \\
\Delta _{8}=\left\langle a_{8}+n_{3}+n_{12}\right\rangle , \\
\Delta _{9}=\left\langle \frac{D}{2}+n_{3}-n_{13}+n_{11}+n_{12}\right\rangle
,%
\end{array}%
\right.
\end{equation}

\begin{equation}
G_{A}\left( a_{9,10}-n_{1},a_{5,6}-n_{3};n_{4}\right) =\dfrac{(-1)^{-\frac{D%
}{2}}}{\Gamma (a_{9}+a_{10}-n_{1})\Gamma (a_{5}+a_{6}-n_{3})}%
\dsum\limits_{n_{14},n_{15}}\phi _{n_{4},n_{14},n_{15}}\;\dfrac{\Delta
_{10}\Delta _{11}\Delta _{12}}{\Gamma (\frac{D}{2}+n_{4})},
\end{equation}

\begin{equation}
\left\{
\begin{array}{l}
\Delta _{10}=\left\langle \frac{D}{2}+n_{4}+n_{14}+n_{15}\right\rangle , \\
\Delta _{11}=\left\langle a_{9}+a_{10}-n_{1}+n_{4}+n_{14}\right\rangle , \\
\Delta _{12}=\left\langle a_{5}+a_{6}-n_{3}+n_{4}+n_{15}\right\rangle ,%
\end{array}%
\right.
\end{equation}

\begin{equation}
\overline{G}_{A}\left( a_{11}+a_{12}-n_{2},-n_{4}\right) =(-1)^{-\frac{D}{2}}%
\dfrac{\Delta _{13}}{\Gamma (a_{11}+a_{12}-n_{2}-n_{4})},
\end{equation}

\begin{equation}
\left\{
\begin{array}{l}
\Delta _{13}=\left\langle a_{11}+a_{12}-n_{2}-n_{4}-\frac{D}{2}\right\rangle
.%
\end{array}%
\right.
\end{equation}%
Finally we obtain the MRE associated to the diagram:

\begin{equation}
\fbox{$%
\begin{array}{ll}
G= & \dfrac{(-1)^{-\frac{5D}{2}}}{\Gamma (a_{7})\Gamma
(a_{8})\prod\nolimits_{j=1}^{4}\Gamma (a_{j})}\dsum\limits_{n_{1},..,n_{15}}%
\phi
_{n_{1},..,n_{17}}(-m_{1}^{2})^{n_{7}}(-m_{2}^{2})^{n_{10}}(-M^{2})^{n_{13}}
\\
&  \\
& \dfrac{1}{\Gamma (\frac{D}{2}+n_{1})\Gamma (\frac{D}{2}+n_{2})\Gamma (%
\frac{D}{2}+n_{3}-n_{13})\Gamma (\frac{D}{2}+n_{4})\Gamma (\frac{D}{2})} \\
&  \\
& \dfrac{\prod\nolimits_{k=1}^{13}\Delta _{k}}{\Gamma
(a_{9}+a_{10}-n_{1})\Gamma (a_{5}+a_{6}-n_{3})\Gamma
(a_{11}+a_{12}-n_{2}-n_{4})}%
\end{array}%
$}
\end{equation}%
Table III shows again a comparison of the IBFE applied modularly or complete
to the diagram.

\begin{equation}
\begin{tabular}{lll}
\hline
& Complete (better factorization) & Modular \\ \hline
Multiplicity multiregion series $\left( {\sigma }\right) $ &
\multicolumn{1}{c}{21} & \multicolumn{1}{c}{15} \\
Kronecker deltas of the expansion $\left( {\delta }\right) $ &
\multicolumn{1}{c}{19} & \multicolumn{1}{c}{13} \\
Multiplicity resulting series $\left( {\sigma -\delta }\right) $ &
\multicolumn{1}{c}{2} & \multicolumn{1}{c}{2} \\
Possible contributions to the solution $\left( {C}_{\delta }^{\sigma
}\right) $ & \multicolumn{1}{c}{210} & \multicolumn{1}{c}{105} \\ \hline
\end{tabular}
\tag{$Table\ III$}
\end{equation}

\section{Comments}

\subsection{Other $1$-$loop$ functions associated to the bubble module}

The $1$-$loop$ functions for diagrams that contain the bubble type
insertions previously deduced do not contemplate all the possible cases. To
see this let us try to solve the following topology:

\begin{equation}
\begin{minipage}{5.5cm} \includegraphics[scale=.7]
{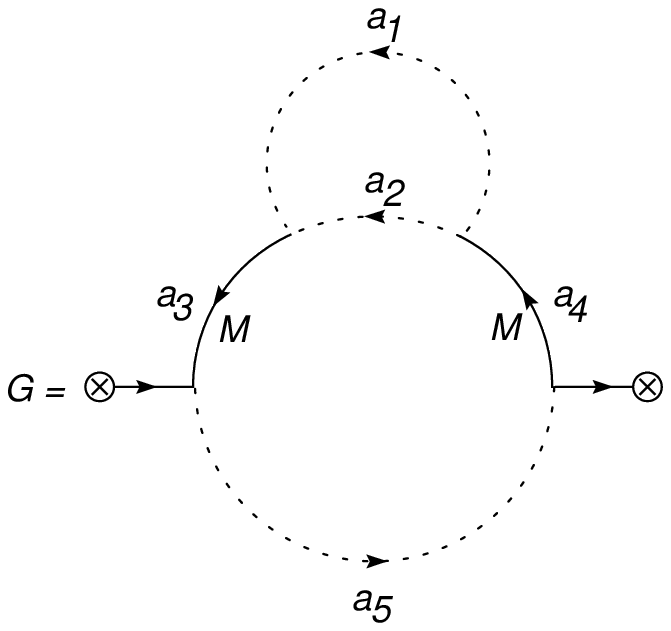} \end{minipage}  \label{ff15}
\end{equation}%
This diagram contains two massive propagators, each of them characterized by
mass $M$, arranged as shown in the figure (continuous line), and the rest of
the propagators are massless (dashed lines). It can be easily evaluated if
Schwinger's parametric representation of the whole diagram is integrated.
Let us see:

\begin{equation}
G=\dfrac{(-1)^{-D}}{\prod\limits_{j=1}^{5}\Gamma (a_{j})}\dint\limits_{0}^{%
\infty }d\overrightarrow{x}\;\frac{\exp (\left( x_{3}+x_{4}\right)
M^{2})\exp \left( -\dfrac{\left[ x_{1}x_{2}+\left( x_{1}+x_{2}\right) \left(
x_{3}+x_{4}\right) \right] x_{5}}{x_{1}x_{2}+\left( x_{1}+x_{2}\right)
\left( x_{3}+x_{4}\right) +x_{5}\left( x_{3}+x_{4}\right) }p^{2}\right) }{%
\left[ x_{1}x_{2}+\left( x_{1}+x_{2}\right) \left( x_{3}+x_{4}\right)
+x_{5}\left( x_{3}+x_{4}\right) \right] ^{\frac{D}{2}}}.
\end{equation}%
The integral is written in a factorized form and it is readily evaluated
with IBFE. After doing the corresponding expansions we realize that the
obtained MRE has ten summations and nine deltas, so the solutions will
correspond to one variable hypergeometric functions $\left(
\;_{q}F_{q-1}\right) $, which is something expected given the type of
topology and the number of energy scales present in the problem. Moreover,
we will have at most ten possible contributions that distribute themselves
in the kinematical regions $\left\vert \dfrac{M^{2}}{p^{2}}\right\vert <1$
and $\left\vert \dfrac{p^{2}}{M^{2}}\right\vert <1$.

Now we will try to solve this diagram, but this time applying the $1$-$loop$
functions deduced in this chapter. The upper loop is easily reduced using
the $1$-$loop$ function $G_{A}$, getting then the following reduced diagram:

\begin{equation}
G=\dsum\limits_{n}G_{A}\left( a_{1},a_{2};n\right) \;\times
\begin{minipage}{4.5cm} \includegraphics[scale=.7]
{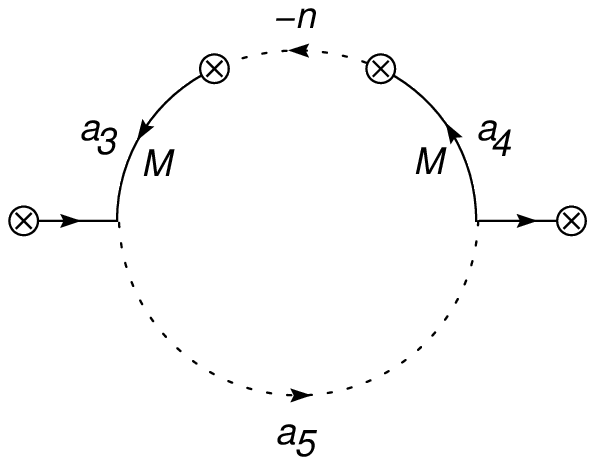} \end{minipage}
\end{equation}%
Nevertheless, we obtain one loop which is not possible to reduce with the $1$%
-$loop$ functions previously defined. The reason is the presence of the
mixed massive propagator in the upper branch of the diagram. Thus it is
necessary to consider a new loop function that we will call $G_{E}$, and
which is associated to the generic topology of the form:

\begin{equation}
\begin{minipage}{5.5cm} \includegraphics[scale=.7]
{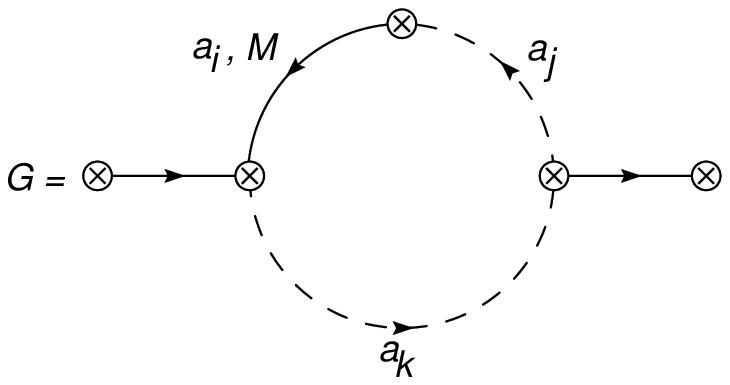} \end{minipage}
\end{equation}%
Let us try to find the MRE for this new loop function. For this we start
with the corresponding Schwinger parametric representation.

\begin{equation}
G=\dfrac{(-1)^{-\frac{D}{2}}}{\Gamma (a_{i})\Gamma (a_{j})\Gamma (a_{k})}%
\dint\limits_{0}^{\infty }d\overrightarrow{x}\;\frac{\exp \left(
x_{i}M^{2}\right) \exp \left( -\dfrac{\left( x_{i}+x_{j}\right) x_{k}}{%
x_{i}+x_{j}+x_{k}}p^{2}\right) }{\left( x_{i}+x_{j}+x_{k}\right) ^{\frac{D}{2%
}}}.
\end{equation}%
When expanding the exponential that contains $p^{2}$ the following series is
obtained:

\begin{equation}
G=\dsum\limits_{n}G_{E}\left( a_{i},a_{j},a_{k};n;M^{2}\right) \;\frac{1}{%
\left( p^{2}\right) ^{-n}},
\end{equation}%
and just as before the $1$-$loop$ function is defined:

\begin{equation}
G_{E}\left( a_{i},a_{j},a_{k};n;M^{2}\right) =\dfrac{(-1)^{-\frac{D}{2}}}{%
\Gamma (a_{i})\Gamma (a_{j})\Gamma (a_{k})}\phi _{n}\dint\limits_{0}^{\infty
}d\overrightarrow{x}\;\frac{\exp \left( x_{i}M^{2}\right) x_{k}^{n}\left(
x_{i}+x_{j}\right) ^{n}}{\left( x_{i}+x_{j}+x_{k}\right) ^{\frac{D}{2}+n}}.
\end{equation}%
Taking into account the complete expansion procedure finally one finds that
the multiregion series for this new $1$-$loop$ function is:

\begin{equation}
\fbox{$G_{E}\left( a_{i},a_{j},a_{k};n;M^{2}\right) =\dfrac{(-1)^{-\frac{D}{2%
}}}{\Gamma (a_{i})\Gamma (a_{j})\Gamma (a_{k})}\dsum\limits_{s_{1},..,s_{5}}%
\phi _{n,s_{1},..,s_{5}}\;(-M_{j}^{2})^{s_{1}}\dfrac{\Delta _{1}...\Delta
_{5}}{\Gamma (\frac{D}{2}+n)\Gamma (-n-s_{2})}$}
\end{equation}%
with the constraints:

\begin{equation}
\left\{
\begin{array}{l}
\Delta _{1}=\left\langle \tfrac{D}{2}+n+s_{2}+s_{3}\right\rangle , \\
\Delta _{2}=\left\langle -n-s_{2}+s_{4}+s_{5}\right\rangle , \\
\Delta _{3}=\left\langle a_{i}+s_{1}+s_{4}\right\rangle , \\
\Delta _{4}=\left\langle a_{j}+s_{5}\right\rangle , \\
\Delta _{5}=\left\langle a_{k}+n+s_{3}\right\rangle .%
\end{array}%
\right.
\end{equation}%
Graphically this result can be represented as follows:

\begin{equation}
\begin{minipage}{5.2cm} \includegraphics[scale=.7]
{caso_comentario3.eps} \end{minipage}%
\begin{array}{cc}
& -n \\
=\dsum\limits_{n}G_{E}\left( a_{i},a_{j},a_{k};n;M^{2}\right) \; & \times
\begin{minipage}{3.3cm} \includegraphics[scale=.8]
{graph_D.eps} \end{minipage} \\
&
\end{array}
\label{ff16}
\end{equation}%
This loop contributes with 6$\Sigma $ and 5$\delta $. Coming back now to our
initial problem presented in equation $\left( \ref{ff15}\right) $, and using
formula $\left( \ref{ff16}\right) $ we find the MRE for this problem:

\begin{equation}
G=\sum\limits_{n,l}G_{A}\left( a_{1},a_{2};n\right) \times G_{E}\left(
a_{3}+a_{4},-n,a_{5};l;M^{2}\right) \;\left( p^{2}\right) ^{l},
\end{equation}%
which can be compared with the equivalent MRE coming from the parametric
representation of the complete diagram (Table IV):

\begin{equation}
\begin{tabular}{lll}
\hline
& Complete (better factorization) & Modular \\ \hline
Multiplicity multiregion series $\left( {\sigma }\right) $ &
\multicolumn{1}{c}{10} & \multicolumn{1}{c}{9} \\
Kronecker deltas of the expansion $\left( {\delta }\right) $ &
\multicolumn{1}{c}{9} & \multicolumn{1}{c}{8} \\
Multiplicity resultanting series $\left( {\sigma -\delta }\right) $ &
\multicolumn{1}{c}{1} & \multicolumn{1}{c}{1} \\
Possible contributions to the solution $\left( {C}_{\delta }^{\sigma
}\right) $ & \multicolumn{1}{c}{10} & \multicolumn{1}{c}{9} \\ \hline
\end{tabular}
\tag{$Table\ IV$}
\end{equation}%
The first we should notice is that in this case once again the expression
for the MRE obtained in terms of modules is more reduced than the similar
expression which evaluates the complete diagram. A second aspect is that
just as the loop function $G_{E}$, there are many others which are
variations of the bubble module and that have not been included here,
although as has been done above any other configuration has an MRE which can
be trivially found.

\subsection{Generic diagrams with bubble type insertions}

We have shown that the IBFE technique is very useful and simple to apply in
diagrams that are built by successive one loop insertions, which can be
either massless or massive. Nevertheless, the modular reduction of a diagram
using the IBFE technique is not only readily applicable to this type of
topologies, but as will be discussed later, the previously deduced formulae
are also useful for reducing any bubble type subgraph that is contained
within a generic graph. As an example let us analyze the following diagram:

\begin{equation}
G=%
\begin{minipage}{3.8cm} \includegraphics[scale=.7]
{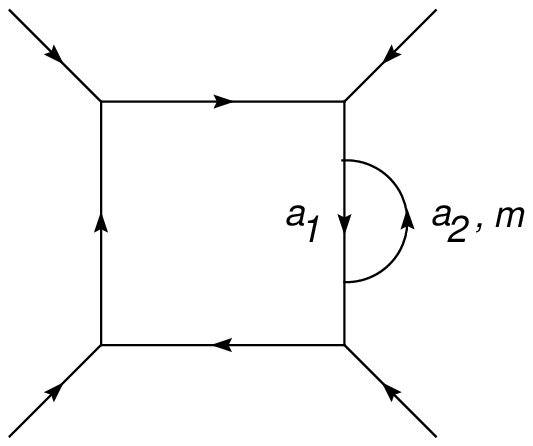} \end{minipage}
\end{equation}%
Independently of the kinematic characteristics of this diagram, our interest
is in the propagator correction with a bubble type insertion and with one of
the propagators with mass $m$. Using the respective loop function in order
to reduce the loop we obtain the following graphical equation:

\begin{equation}
G=\sum\limits_{n}G_{B}(a_{2},a_{1};n;m^{2})\;\times
\begin{minipage}{3.1cm} \includegraphics[scale=.7]
{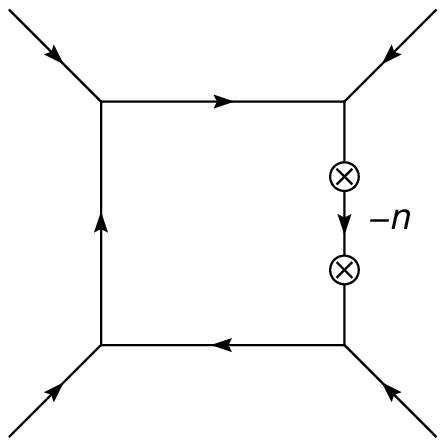} \end{minipage}
\end{equation}%
From a topological point of view one can see that there is a reduction, so
to find the MRE of this resulting one loop diagram is simpler than the
original two loop one. Just as in this example, it is possible to use the $1$%
-$loop$ functions in generic diagrams whose propagators contain bubble type
corrections and its variations, which simplifies considerably the search of
an MRE of certain diagrams and its solutions.

\subsection{$n$-$loop$ modules and $n$-$loop$ functions}

We have shown that it is possible to apply IBFE to loop by loop recursively
built subgraphs and find the respective MRE in terms of the $1$-$loop$
functions defined above. Nevertheless, the flexibility of IBFE goes beyond
that, and it is possible to generate also $n$-$loop$ functions or
equivalently module functions that consider two or more loops
simultaneously. Such a module, with $n$ loops, will lead to a growing number
of $n$-$loop$ functions when one takes into account the different
possibilities of distributing the propagator masses of the $n$-$loop$
module. In general to work modularly is advantageous for reducing diagrams,
since it simplifies the search of the MRE. Nevertheless, if the idea is to
have less algebraic manipulations and a minimal expression for the MRE of
the topology, then the previous examples indicate that the MRE is going to
be minimal when the modules under consideration are compositions of only one
loop.

\section{Conclusions}

\qquad In this work we have described an efficient and easy way to to
implement the integration technique IBFE, applied to a specific Feynman
diagram. The result that has been reached using the concept of modular
reduction has two important characteristics: first, in general the diagram
multiregion expansion of a given diagram is shorter than if the MRE is
obtained from the parametric representation of the complete diagram; second,
the modular treatment operationally facilitates obtaining the multiregion
expansion. Let us discuss this last point, comparing the modular reduction
in cases of loop by loop reducible diagrams with $1$-$loop$ insertions in a
massless theory which are evaluated using conventional calculation. The
difference is that with IBFE it is possible to do a calculation with the
same degree of simplicity, although for more complex bubble type diagrams it
is now possible to include masses.

The modular reduction idea leads in general to a minimization of the MRE of
a diagram, which is more evident when the modules to reduce are composed of
one loop. In fact, the loop by loop reduction optimizes the IBFE
application. In more general terms and beyond this work, depending on the
one loop topology the minimal MRE is reached by a systematic reduction going
from the topologically simplest to the more complex: bubbles, Triangles,
Boxes, etc.

The importance of a minimal MRE is of course tied to a minimization of the
time of finding the solutions, since the MRE gets the solutions to the
parametric integrals evaluating summations with Kronecker deltas in all
possible combinations, and now the MRE is shorter (less summations and
Kronecker deltas, which diminish equally), then there are less combinations
that need to be taken into account in order to find the solution. Basically
what happens is that several combinations which do not lead to relevant
terms in the solution are eliminated, and given that with IBFE the
integration is replaced by a linear system evaluation, the minimized MRE
implies smaller linear systems, and therefore simpler to analyze.

The integration technique IBFE is a simple method and at the same time
powerful, which can be advantageously compared with other Feynman diagram
evaluation, and which does not require great mathematical knowledge.
Although here we have concentrated in a scalar theory, this method is
directly applicable to other theories by previous scalarization of the
tensorial loop integrals, which produces a sum of scalar integrals. In
general the most complicated case of an $N$ propagator tensorial integral is
precisely the corresponding $N$ propagator scalar integral. This is the
reason for improving or creating numerical or analytical techniques for the
evaluation of scalar integrals in perturbative studies in field theory.

\bigskip

\subparagraph{Acknowledgements :}

We acknowledge support from Fondecyt (Chile) under Grant No. 3080029 and
Grant Fund. Andes No. C-14055/20.

\newpage

\appendix

\section{Summary of formulae for the bubble module}

\begin{equation*}
\fbox{$%
\begin{array}{l}
\begin{array}{cc}
a_{j} &  \\
\begin{minipage}{4.5cm} \includegraphics[scale=.7]
{graph_A.eps} \end{minipage} &
\begin{array}{cc}
& -n \\
=\dsum\limits_{n}G_{A}\left( a_{j},a_{k};n\right) & \times
\begin{minipage}{3.6cm} \includegraphics[scale=.8]
{graph_D.eps} \end{minipage} \\
&
\end{array}
\\
a_{k} &
\end{array}
\\
Composition\Longrightarrow 3\Sigma /3\delta . \\
\\
\\
\begin{array}{cc}
a_{j},M_{j} &  \\
\begin{minipage}{4.5cm} \includegraphics[scale=.7]
{graph_A.eps} \end{minipage} &
\begin{array}{cc}
& -n \\
=\dsum\limits_{n}G_{B}(a_{j},a_{k};n;M_{j}^{2}) & \times
\begin{minipage}{3.6cm} \includegraphics[scale=.8]
{graph_D.eps} \end{minipage} \\
&
\end{array}
\\
a_{k} &
\end{array}
\\
Composition\Longrightarrow 4\Sigma /3\delta . \\
\\
\\
\begin{array}{cc}
a_{j},M &  \\
\begin{minipage}{4.5cm} \includegraphics[scale=.7]
{graph_A.eps} \end{minipage} &
\begin{array}{cc}
& -n \\
=\dsum\limits_{n}G_{C}\left( a_{j},a_{k};n;M^{2}\right) & \times
\begin{minipage}{3.6cm} \includegraphics[scale=.8]
{graph_D.eps} \end{minipage} \\
&
\end{array}
\\
a_{k},M &
\end{array}
\\
Composition\Longrightarrow 4\Sigma /3\delta . \\
\\
\\
\begin{array}{cc}
a_{j},M_{j} &  \\
\begin{minipage}{4.5cm} \includegraphics[scale=.7]
{graph_A.eps} \end{minipage} &
\begin{array}{cc}
& -n \\
=\dsum\limits_{n}G_{D}(a_{j},a_{k};n;M_{j}^{2},M_{k}^{2}) & \times
\begin{minipage}{3.6cm} \includegraphics[scale=.8]
{graph_D.eps} \end{minipage} \\
&
\end{array}
\\
a_{k},M_{k} &
\end{array}
\\
Composition\Longrightarrow 5\Sigma /3\delta .%
\end{array}%
$}
\end{equation*}

\newpage

\begin{equation*}
\fbox{$%
\begin{array}{l}
\begin{array}{cc}
a_{j} &  \\
\begin{minipage}{2.5cm} \includegraphics[scale=.7]
{graph_B.eps} \end{minipage} & =\overline{G}_{A}\left( a_{j},a_{k}\right) \\
a_{k} &
\end{array}
\\
Composition\Longrightarrow 0\Sigma /1\delta . \\
\\
\\
\begin{array}{cc}
a_{j},M_{j} &  \\
\begin{minipage}{2.5cm} \includegraphics[scale=.7]
{graph_B.eps} \end{minipage} & =\overline{G}_{B}(a_{j},a_{k};M_{j}^{2}) \\
a_{k} &
\end{array}
\\
Composition\Longrightarrow 2\Sigma /2\delta . \\
\begin{array}{l}
\\
\\
\begin{array}{cc}
a_{j},M &  \\
\begin{minipage}{2.5cm} \includegraphics[scale=.7]
{graph_B.eps} \end{minipage} & =\overline{G}_{C}\left(
a_{j},a_{k};M^{2}\right) \\
a_{k},M &
\end{array}
\\
Composition\Longrightarrow 2\Sigma /2\delta . \\
\\
\\
\begin{array}{cc}
a_{j},M_{j} &  \\
\begin{minipage}{2.5cm} \includegraphics[scale=.7]
{graph_B.eps} \end{minipage} & =\overline{G}%
_{D}(a_{j},a_{k};M_{j}^{2},M_{k}^{2}) \\
a_{k},M_{k} &
\end{array}
\\
Composition\Longrightarrow 4\Sigma /3\delta .%
\end{array}%
\end{array}%
$}
\end{equation*}

\section{Hypergeometric functions}

\qquad In the evaluation of $L$ loop Feynman diagrams, all the possible
solutions that can be found are presented in terms of multivariable
generalized hypergeometric series. The actual value of the variables is
given in general by ratios of two energy scales associated to the graph \cite%
{IGoIBFE}, including also arguments of value '1'. The purpose of this
appendix is to provide the necessary information about simple and double
hypergeometric functions \cite{WBa, GGa, LSl, LGr, HEx}, which appear in the
solution of the loop integrals, and in particular present their convergence
conditions, which are what finally define the different kinematical regions
associated to the solution of a specific Feynman diagram.

\subsection{Definition of the generalized hypergeometric function}

The functions associated to one variable series solutions of a Feynman
diagram are always expressible in terms of generalized hypergeometric
functions, in particular the hypergeometric function of order $\left(
q,q-1\right) $, which is the only one variable series that is solution to
the loop integrals. Thsi function is conventionally denoted as:

\begin{equation}
\;_{q}F_{q-1}(\left\{ a\right\} ;\left\{ b\right\} ;z)\equiv
\;_{q}F_{q-1}\left( \left.
\begin{array}{c}
\left\{ a\right\} \\
\left\{ b\right\}%
\end{array}%
\right\vert z\right) \equiv \dsum\limits_{k=0}^{\infty }\frac{\left(
a_{1}\right) _{k}...\left( a_{q}\right) _{k}}{\left( b_{1}\right)
_{k}...\left( b_{q-1}\right) _{k}}\frac{z^{k}}{k!},
\end{equation}%
where the factors $\left( \alpha \right) _{k}$ are called Pochhammer symbols
and which are defined by:

\begin{equation}
\left( \alpha \right) _{k}=\dfrac{\Gamma (\alpha +k)}{\Gamma (\alpha )}.
\label{ff34}
\end{equation}%
For these functions the convergence conditions are:

\begin{itemize}
\item The generalized hypergeometric functions are given by series defined
in the convergence region $\left\vert z\right\vert <1$, while for $%
\left\vert z\right\vert >1$ they are defined by analytical continuation.

\item If $\left\vert z\right\vert <1$ the series converges absolutely. Since
the variable $z$ represents the ratio between two energy scales of the
topology, what is obtained is one of the limit representations or solutions,
an expansion around $z\rightarrow 0$.

\item If $z=1$, the necessary requirement for the series convergence is that
$\Re e\left( \omega \right) >0$, where $\omega $ is called parametric excess
and it is given by the equation:
\end{itemize}

\begin{equation}
\omega =\sum\limits_{j=0}^{q}b_{j}-\sum\limits_{j=0}^{q+1}a_{j}.
\end{equation}

\begin{itemize}
\item For the convergence in $z=-1$ it is sufficient that $\Re e\left(
\omega \right) >-1$.
\end{itemize}

\subsection{Some identities of the Pochhammer symbols}

The following identities are very useful for building the hypergeometric
function starting from the contributions that are obtained of the MRE of an
arbitrary diagram $G$. Aside from formula $\left( \ref{ff34}\right) $,
sometimes it is useful to use other identities that are needed when there
are factors of the type $\Gamma (a\pm n)$ and $\Gamma (a\pm 2n)$ in the
solutions, such as:

\begin{equation}
\left( a\right) _{-n}=\dfrac{\Gamma (a-n)}{\Gamma (a)}=\dfrac{(-1)^{n}}{%
(1-a)_{n}},
\end{equation}

\begin{equation}
(a)_{2n}=\dfrac{\Gamma (a+2n)}{\Gamma (a)}=4^{n}\left( \dfrac{a}{2}\right)
_{n}\left( \dfrac{a}{2}+\dfrac{1}{2}\right) _{n}.
\end{equation}

\subsection{Two variable hypergeometric functions}

Here we describe two variable hypergeometric functions, which correspond to
the Kamp\'{e} de F\'{e}riet generalized double hypergeometric function $F
^{\substack{ p:r:u  \\ q:s:v}}$ and the generalized hypergeometric $%
\overline{F}^{\substack{ p:r:u  \\ q:s:v}}$.

\subsubsection{Funci\'{o}n Kamp\'{e} de F\'{e}riet $F^{\protect\substack{ %
p:r:u  \\ q:s:v}}$}

This function is defined as:

\begin{equation}
\begin{array}{ll}
F^{\substack{ p:r:u  \\ q:s:v}}\left( \left.
\begin{array}{ccc}
\alpha _{1},...,\alpha _{p} & a_{1},...,a_{r} & c_{1},...,c_{u} \\
\beta _{1},...,\beta _{q} & b_{1},...,b_{s} & d_{1},...,d_{v}%
\end{array}%
\right\vert x,y\right) & \equiv F^{\substack{ p:r:u  \\ q:s:v}}\left( \left.
\begin{array}{ccc}
\{\alpha \} & \{a\} & \{c\} \\
\{\beta \} & \{b\} & \{d\}%
\end{array}%
\right\vert x,y\right) \\
&  \\
& \equiv \dsum\limits_{n,m}^{\infty }\dfrac{\prod\limits_{j=1}^{p}(\alpha
_{j})_{n+m}\prod\limits_{j=1}^{r}(a_{j})_{n}\prod\limits_{j=1}^{u}(c_{j})_{m}%
}{\prod\limits_{j=1}^{q}(\beta
_{j})_{n+m}\prod\limits_{j=1}^{s}(b_{j})_{n}\prod\limits_{j=1}^{v}(d_{j})_{m}%
}\dfrac{x^{n}}{n!}\dfrac{y^{m}}{m!},%
\end{array}%
\end{equation}%
where the convergence conditions of the double series exist if the following
relation between the indices are satisfied:

\begin{equation}
p+r\leqslant q+s+1,
\end{equation}

\begin{equation}
p+u\leqslant q+v+1,
\end{equation}%
and if furthermore, the arguments fulfil the condition:

\begin{equation*}
\begin{array}{lll}
\left\vert x\right\vert ^{\tfrac{1}{(p-q)}}+\left\vert y\right\vert ^{\tfrac{%
1}{(p-q)}}<1 &  & \text{, if }(p>q), \\
&  &  \\
\max \{\left\vert x\right\vert ,\left\vert y\right\vert \}<1 &  & \text{, if
}(p\leqslant q).%
\end{array}%
\end{equation*}

\subsubsection{Function $\overline{F}^{\protect\substack{ p:r:u  \\ q:s:v }}
$}

This series, which appears frequently in the solutions to Feynman diagrams,
has the following definition:

\begin{equation}
\begin{array}{ll}
\overline{F}^{\substack{ p:r:u  \\ q:s:v}}\left( \left.
\begin{array}{ccc}
\alpha _{1},...,\alpha _{p} & a_{1},...,a_{r} & c_{1},...,c_{u} \\
\beta _{1},...,\beta _{q} & b_{1},...,b_{s} & d_{1},...,d_{v}%
\end{array}%
\right\vert x,y\right) & =\overline{F}^{\substack{ p:r:u  \\ q:s:v}}\left(
\left.
\begin{array}{ccc}
\{\alpha \} & \{a\} & \{c\} \\
\{\beta \} & \{b\} & \{d\}%
\end{array}%
\right\vert x,y\right) \\
&  \\
& =\dsum\limits_{n,m}^{\infty }\dfrac{\tprod\limits_{j=1}^{p}(\alpha
_{j})_{n-m}\prod\limits_{j=1}^{r}(a_{j})_{n}\prod\limits_{j=1}^{u}(c_{j})_{m}%
}{\prod\limits_{j=1}^{q}(\beta
_{j})_{n-m}\prod\limits_{j=1}^{s}(b_{j})_{n}\prod\limits_{j=1}^{v}(d_{j})_{m}%
}\dfrac{x^{n}}{n!}\dfrac{y^{m}}{m!},%
\end{array}%
\end{equation}%
where the convergence conditions of the double series exist if the following
relation between the indices are satisfied:

\begin{equation}
p+r\leqslant q+s+1,
\end{equation}

\begin{equation}
q+u\leqslant p+v+1.
\end{equation}%
All the series that have been found with the technique used here fulfil this
indexes condition. The determination of the convergence region of the
variables can be done using the Horns general convergence theory \cite{HEx}.

\section{Mathematical Formalism of the integration by fractional expansion
Method}

\subsection{Introduction}

In section $\left( 2.2\right) $ we already introduced some algebraic aspects
of the IBFE technique, which are generated from the identity associated to
the integral parametrization of loops and which is known as Schwinger's
parametrization:

\begin{equation}
\frac{1}{A^{\beta }}=\frac{1}{\Gamma (\beta )}\int\limits_{0}^{\infty
}dx\;x^{\beta -1}\exp (-Ax).  \label{f1}
\end{equation}%
It is possible to find an operational equivalence between the integral
symbol and a Kronecker delta, given by:

\begin{equation}
\int dx\;x^{\beta +n-1}\equiv \Gamma \left( \beta \right) \dfrac{\Gamma
\left( n+1\right) }{\left( -1\right) ^{n}}\;\delta _{\beta +n,0}.  \label{f2}
\end{equation}%
For simplicity we have eliminated the integral limits, since this identity
has only validity in the context of the integrand expansion in $\left( \ref%
{f1}\right) $. This expression is crucial for the development of the IBFE
method, since in the Feynman diagram evaluation the corresponding Schwinger
parametric representation is a generalized structure of the expression $%
\left( \ref{f1}\right) $.

\subsubsection{Some Properties}

For the study of some properties of $\left( \ref{f2}\right) $ it is
convenient to use the notation defined in $\left( \ref{f21}\right) $ in
order to help us formalize the mechanism of the IBFE technique. Then let:

\begin{equation}
\int dx\;x^{\nu _{1}+\nu _{2}-1}\equiv \left\langle \nu _{1}+\nu
_{2}\right\rangle ,  \label{f4}
\end{equation}%
where $\nu _{1}$ and $\nu _{2}$ are indices which can take arbitrary values.

\paragraph{Property l. Commutativity of indices :}

We can explicitly write $\left( \ref{f4}\right) $ in two possible forms
according to formula $\left( \ref{f2}\right) $:

\begin{equation}
\begin{array}{cc}
\left\langle \nu _{1}+\nu _{2}\right\rangle = & \left\{
\begin{array}{c}
\Gamma \left( \nu _{1}\right) \dfrac{\Gamma \left( \nu _{2}+1\right) }{%
\left( -1\right) ^{\nu _{2}}}\;\delta _{\nu _{1}+\nu _{2},0}, \\
\\
\Gamma \left( \nu _{2}\right) \dfrac{\Gamma \left( \nu _{1}+1\right) }{%
\left( -1\right) ^{\nu _{1}}}\;\delta _{\nu _{1}+\nu _{2},0}.%
\end{array}%
\right.%
\end{array}
\label{f10}
\end{equation}%
Starting from $\left( \ref{f1}\right) $ one can show the equivalence of both
forms in $\left( \ref{f10}\right) $, and for this it is enough to expand the
integrand exponential and replace $\left\langle \cdot \right\rangle $ for
each case:

Let us consider the following integral representation:

\begin{equation}
\frac{1}{A^{\nu _{1}}}=\frac{1}{\Gamma \left( \nu _{1}\right) }%
\int\limits_{0}^{\infty }dx\;x^{\nu _{1}-1}\exp \left( -Ax\right) ,
\label{f5}
\end{equation}%
which in terms of a series, in the proposed sense $\left( \ref{f4} \right) $%
, turns out to be:

\begin{equation}
\frac{1}{A^{\nu _{1}}}=\frac{1}{\Gamma \left( \nu _{1}\right) }%
\sum\limits_{\nu _{2}}\frac{\left( -1\right) ^{\nu _{2}}}{\Gamma \left( \nu
_{2}+1\right) }A^{\nu _{2}}\left\langle \nu _{1}+\nu _{2}\right\rangle .
\end{equation}%
Selecting now $\left\langle \nu _{1}+\nu _{2}\right\rangle =\Gamma \left(
\nu _{1}\right) \dfrac{\Gamma \left( \nu _{2}+1\right) }{\left( -1\right)
^{\nu _{2}}}\;\delta _{\nu _{1}+\nu _{2},0}$, we directly obtain the
equality in $\left( \ref{f5}\right) $. Analogously selecting $\left\langle
\nu _{1}+\nu _{2}\right\rangle =\Gamma \left( \nu _{2}\right) \dfrac{\Gamma
\left( \nu _{1}+1\right) }{\left( -1\right) ^{\nu _{1}}}\;\delta _{\nu
_{1}+\nu _{2},0}$ and using then the identity:

\begin{equation}
\dfrac{\Gamma \left( y\right) }{\Gamma \left( y-z\right) }=\left( -1\right)
^{-z}\dfrac{\Gamma \left( 1+z-y\right) }{\Gamma \left( 1-y\right) },
\end{equation}%
with $y=\nu _{1}$ and $z=2\nu _{1}$, the equality $\left( \ref{f5}\right) $
is finally obtained, which shows the equivalence between the two ways of
writing $\left\langle \nu _{1}+\nu _{2}\right\rangle $.

Another way of writing $\left( \ref{f4}\right) $, which is useful for the
simplification of terms that contain the factor:

\begin{equation}
\frac{(-1)^{m}}{\Gamma (m+1)},
\end{equation}%
where $m$ is an arbitrary index, is the following:

\begin{equation}
\left\langle \nu _{1}+\nu _{2}\right\rangle =\left\langle \nu _{1}+\nu
_{2}-m+m\right\rangle =\left\langle -m+m\right\rangle =\Gamma (-m)\frac{%
\Gamma (m+1)}{(-1)^{m}}\;\delta _{\nu _{1}+\nu _{2},0}.
\end{equation}%
Notice that explicit use of the Kronecker deltas has been made, in order to
simplify the parenthesis $\left\langle \cdot \right\rangle $, and since this
happens in the context of expansions, the Kronecker deltas remain to
indicate the constraints between the indices $\nu _{1}$ and $\nu _{2}$.

\paragraph{Property ll. Significance of the Multiregion Expansion MRE :}

Let us consider the following binomial expansion:

\begin{equation}
\left( A_{1}+A_{2}\right) ^{\pm \nu },  \label{f13}
\end{equation}%
where the quantities $A_{1}$,$A_{2}$ and $\nu $ can take arbitrary values.
In this case there are two possible regions or limits for the expansion: the
region where $\left( A_{1}>A_{2}\right) $ and the region where $\left(
A_{1}<A_{2}\right) $. These expansions are respectively:

\begin{enumerate}
\item Regi\'{o}n $\left( A_{1}>A_{2}\right) $
\end{enumerate}

\begin{equation}
\left( A_{1}+A_{2}\right) ^{\pm \nu }=A_{1}^{\pm \nu
}\sum\limits_{n=0}^{\infty }\dfrac{\left( \mp \nu \right) _{n}}{\Gamma (n+1)}%
\left( -\dfrac{A_{2}}{A_{1}}\right) ^{n}.  \label{f15}
\end{equation}

\begin{enumerate}
\item[2.] Regi\'{o}n $\left( A_{1}<A_{2}\right) $
\end{enumerate}

\begin{equation}
\left( A_{1}+A_{2}\right) ^{\pm \nu }=A_{2}^{\pm \nu
}\sum\limits_{n=0}^{\infty }\dfrac{\left( \mp \nu \right) _{n}}{\Gamma (n+1)}%
\left( -\dfrac{A_{1}}{A_{2}}\right) ^{n}.  \label{f14}
\end{equation}%
The factor $(\nu )_{n}$ is the Pochhammer symbol and it is given by:

\begin{equation}
(\nu )_{n}=\dfrac{\Gamma \left( \nu +n\right) }{\Gamma \left( \nu \right) }.
\end{equation}%
We have obtained in this manner expansions in the two possible limits
separately. It is possible, however, to express both results employing a
single series which contains simultaneously both regions. In this sense we
can say that this type of expansion corresponds to a multiregion series
representation of the binomial. To show this, let us express the binomial $%
\left( \ref{f13}\right) $ using the integral representation of the
denominator indicated in $\left( \ref{f1}\right) $. Then we get:

\begin{equation}
\left( A_{1}+A_{2}\right) ^{\pm \nu }=\frac{1}{\Gamma (\mp \nu )}%
\int\limits_{0}^{\infty }dx\;x^{\mp \nu -1}\exp (-xA_{1})\exp (-xA_{2}),
\end{equation}%
and the exponentials are expanded separately, obtaining:

\begin{equation}
\left( A_{1}+A_{2}\right) ^{\pm \nu }=\frac{1}{\Gamma (\mp \nu )}%
\sum\limits_{n_{1}}\sum\limits_{n_{2}}\frac{(-1)^{n_{1}+n_{2}}}{\Gamma
(n_{1}+1)\Gamma (n_{2}+1)}A_{1}^{n_{1}}A_{2}^{n_{2}}\int dx\;x^{\mp \nu
+n_{1}+n_{2}-1}.
\end{equation}%
Using the identity $\left( \ref{f4}\right) $ we get the multiregion binomial
expansion:

\begin{equation}
\left( A_{1}+A_{2}\right) ^{\pm \nu }=\frac{1}{\Gamma (\mp \nu )}%
\sum\limits_{n_{1}}\sum\limits_{n_{2}}\dfrac{(-1)^{n_{1}+n_{2}}}{\Gamma
(n_{1}+1)\Gamma (n_{2}+1)}A_{1}^{n_{1}}A_{2}^{n_{2}}\left\langle \mp \nu
+n_{1}+n_{2}\right\rangle ,  \label{f8}
\end{equation}%
where according to property $\left( \ref{f10}\right) $ we can express the
parenthesis $\left\langle \cdot \right\rangle $ in three different ways. In
any of them we will have the same Kronecker delta which eliminates one of
the two sums. Let us see:

\begin{equation}
\begin{array}{cc}
\left\langle \mp \nu +n_{1}+n_{2}\right\rangle = & \left\{
\begin{array}{l}
\Gamma \left( \mp \nu +n_{1}\right) \dfrac{\Gamma \left( n_{2}+1\right) }{%
\left( -1\right) ^{n_{2}}}\;\delta _{\mp \nu +n_{1}+n_{2},0} \\
\\
\Gamma \left( \mp \nu +n_{2}\right) \dfrac{\Gamma \left( n_{1}+1\right) }{%
\left( -1\right) ^{n_{1}}}\;\delta _{\mp \nu +n_{1}+n_{2},0} \\
\\
\Gamma \left( n_{2}+n_{1}\right) \dfrac{\Gamma \left( \mp \nu +1\right) }{%
\left( -1\right) ^{\mp \nu }}\;\delta _{\mp \nu +n_{1}+n_{2},0}.%
\end{array}%
\right.%
\end{array}%
\end{equation}%
On the other hand, the number of possible ways of summing $\left( \ref{f8}%
\right) $ using the Kronecker delta can be found in general by evaluating
the combinatorial $C_{Deltas}^{Summations}$, which in this case is $%
C_{1}^{2}=2$. Let us see what happens when we sum with respect to one
particular index:

\begin{enumerate}
\item \textbf{Sum respect to }$n_{2}$
\end{enumerate}

\qquad Let us use for this case the following equality:

\begin{equation}
\left\langle \mp \nu +n_{1}+n_{2}\right\rangle =\Gamma \left( \mp \nu
+n_{1}\right) \dfrac{\Gamma \left( n_{2}+1\right) }{\left( -1\right) ^{n_{2}}%
}\;\delta _{\mp \nu +n_{1}+n_{2},0},
\end{equation}%
and then replacing in $\left( \ref{f8}\right) $ we obtain:

\begin{equation}
\left( A_{1}+A_{2}\right) ^{\pm \nu }=\frac{1}{\Gamma (\mp \nu )}%
\sum\limits_{n_{1}}(-1)^{n_{1}}\dfrac{\Gamma \left( \mp \nu +n_{1}\right) }{%
\Gamma (n_{1}+1)}A_{1}^{n_{1}}A_{2}^{\pm \nu -n_{1}},
\end{equation}%
or equivalently:

\begin{equation}
\left( A_{1}+A_{2}\right) ^{\pm \nu }=A_{2}^{\pm \nu
}\sum\limits_{n_{1}=0}^{\infty }\dfrac{\left( \mp \nu \right) _{n_{1}}}{%
\Gamma (n_{1}+1)}\left( -\frac{A_{1}}{A_{2}}\right) ^{n_{1}},
\end{equation}%
which gives the expansion associated to the region $\left(
A_{1}<A_{2}\right) $, obtained previously in $\left( \ref{f14}\right) $.

\begin{enumerate}
\item[2.] \textbf{Sum respect to} $n_{1}$
\end{enumerate}

\qquad Analogously, we now use the identity:

\begin{equation}
\left\langle \mp \nu +n_{1}+n_{2}\right\rangle =\Gamma \left( \mp \nu
+n_{2}\right) \dfrac{\Gamma (n_{1}+1)}{\left( -1\right) ^{n_{1}}}\;\delta
_{\mp \nu +n_{1}+n_{2},0},
\end{equation}%
and replacing in $\left( \ref{f8}\right) $ gives:

\begin{equation}
\left( A_{1}+A_{2}\right) ^{\pm \nu }=A_{1}^{\pm \nu
}\sum\limits_{n_{2}=0}^{\infty }\frac{\left( \mp \nu \right) _{n_{2}}}{%
\Gamma (n_{2}+1)}\left( -\frac{A_{2}}{A_{1}}\right) ^{n_{2}},
\end{equation}%
expression that was found in $\left( \ref{f15}\right) $, and valid in the
region $\left( A_{1}>A_{2}\right) $.

The fundamental idea that has been exposed in the previous demonstration is
that using the definition $\left( \ref{f2}\right) $ it is possible to make a
binomial expansion which differs from the conventional on the sense that the
expansion is now around zero and infinity simultaneously. This can be
generalized for multinomials and obtain its MRE:

\begin{equation}
\left( A_{1}+...+A_{l}\right) ^{\pm \nu
}=\sum\limits_{n_{1}}...\sum\limits_{n_{l}}\phi
_{n_{1},..,n_{l}}\;A_{1}^{n_{1}}...A_{l}^{n_{l}}\frac{\left\langle \mp \nu
+n_{1}+...+n_{l}\right\rangle }{\Gamma (\mp \nu )},  \label{f9}
\end{equation}

The number of different expressions that can be extracted starting from
equation $\left( \ref{f9}\right) $ is given by all the possible forms of
evaluating some of the sums, using for this purpose the Kronecker delta
generated by the same expansion, that is $C_{1}^{n_{l}}=n_{l}$ possible
forms. Generalizing even more, any function expressed of a multiregion
series employing $\sigma $ sums and $\delta $ Kronecker deltas, has at most:

\begin{equation}
C_{\delta }^{\sigma }=\dfrac{\sigma !}{\delta !(\sigma -\delta )!}
\end{equation}%
possible ways of being evaluated, and each of these expansions corresponds
to a series of multiplicity $\mu =\left( \sigma -\delta \right) $.

All the resulting series are series representations with respect to the
ratios between the terms of the multinomial and all of them correspond to
multivariable generalizations of the hypergeometric function.

\end{document}